\documentclass[11pt]{article}

\usepackage{amsmath}
\usepackage{amssymb} 
\usepackage{caption}
\usepackage{graphicx}
\usepackage{authblk}
\usepackage[hypertexnames=false,colorlinks=true,linkcolor=blue,citecolor=blue]{hyperref}
\usepackage[numbers,comma,square,sort&compress]{natbib}
\usepackage[a4paper,text={6.5in,10in},centering]{geometry}

\graphicspath{{/}{eps/}{pdf/}{images/}}

\makeatletter\@addtoreset{equation}{section}\makeatother

\newtheorem{Lemma}{Lemma}[section]

\newtheorem{Proposition}[Lemma]{Proposition}

 {\begin{trivlist} \item[]{\bf Proof. }}%
 {\hspace*{\fill}$\rule{.4\baselineskip}{.4\baselineskip}$\end{trivlist}}

 {\begin{trivlist}\item[]\textbf{Acknowledgments.}}{\end{trivlist}}

\newtheorem{conjecture}{Conjecture}

{\begin{trivlist}\item[]\textbf{Proof#1 }}%
 {\hspace*{\fill}$\rule{0.3\baselineskip}{0.35\baselineskip}$\end{trivlist}}

\newcommand{\D}{\mathrm{d}}

\begin{document}

\title{Hexagon Invasion Fronts Outside the Homoclinic Snaking Region in the Planar Swift-Hohenberg Equation}
\author[1]{David J.B. Lloyd}
\affil[1]{\small Department of Mathematics, University of Surrey, Guildford, GU2 7XH, UK}
\date{\today}
\maketitle

\begin{abstract}
Stationary fronts connecting the trivial state and a cellular (distorted) hexagonal pattern in the Swift-Hohenberg equation with a quadratic-cubic nonlinearity are known to undergo a process of infinitely many folds as a parameter is varied, known as homoclinic snaking, where new hexagon cells are added to the core, leading to a region of infinitely-many, co-existing localized states. Outside the homoclinic snaking region, the hexagon fronts can invade the trivial state in a bursting fashion. In this paper, we use a far-field core decomposition to set up a numerical path-following routine to trace out the bifurcation diagrams of hexagon fronts for the two main orientations of cellular hexagon pattern with respect to the interface in the bistable region. We find for one orientation that the hexagon fronts can destabilize as the distorted hexagons are stretched in the transverse direction leading to defects occurring in the deposited cellular pattern. We then plot diagrams showing when the selected fronts for the two main orientations, aligned perpendicular to each other, are compatible leading to a hexagon wavenumber selection prediction for hexagon patches on the plane.
Finally, we verify the compatibility criterion for hexagon patches in the Swift-Hohenberg equation with a quadratic-cubic nonlinearity and a large non-variational perturbation. The numerical algorithms presented in this paper can be adapted to general reaction-diffusion systems.

\end{abstract}

\section{Introduction}
\begin{figure}[h]
\centering
\includegraphics[width=\linewidth]{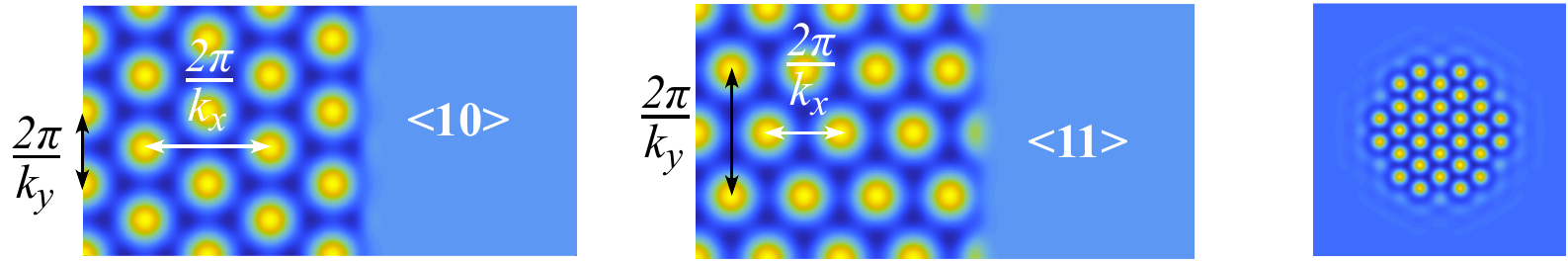}
\caption{The $\langle10\rangle$- and $\langle11\rangle$-hexagon fronts and patches studied in this paper. The hexagon fronts are considered in this paper on the domain that is infinite in $x$ and periodic in $y$ strip while the patches are considered on the infinite plane.\label{f:hex_front_patch}}
\end{figure}

In this paper, we investigate the pattern selection of stationary and invading localized cellular hexagon fronts and patches connecting to the trivial state in the planar Swift-Hohenberg (SH) equation
\begin{equation}\label{e:sh}
u_t = -(1+\Delta)^2u - \mu u + \nu u^2 - u^3,\qquad \mu,\nu\in\mathbb{R},
\end{equation}
where $u = u(x,y,t)$, $\Delta$ is the 2D Laplacian, $\nu\neq0$, and $\mu>0$; see Figure~\ref{f:hex_front_patch}. The SH equation serves as a prototypical model for pattern formation in a range of applications such as magnetizable fluids~\cite{lloyd2015,Groves2017}, vegetation ecology~\cite{zelnik2017,zelnik2018,Tlidi2018,gandhi2019,wuyts2019,bel2012}, phyllotaxis~\cite{pennybacker2015}, fluid systems~\cite{knobloch2015,beaume2018}, nonlinear optics~\cite{odent2016,vladimirov2011}, phase-field crystals~\cite{ophaus2018,Subramanian_2018,emmerich2012}, solidification~\cite{barros2013}, cylinder buckling~\cite{thompson2015,champneys2019} and urban crime~\cite{lloyd2013} to name but a few. There has also been a growing interest in revisiting old problems of front invasion~\cite{castillo2019,Alvarez2019,archer2012} typically into the unstable state. 

\begin{figure}[h]
\centering
\includegraphics[width=\linewidth]{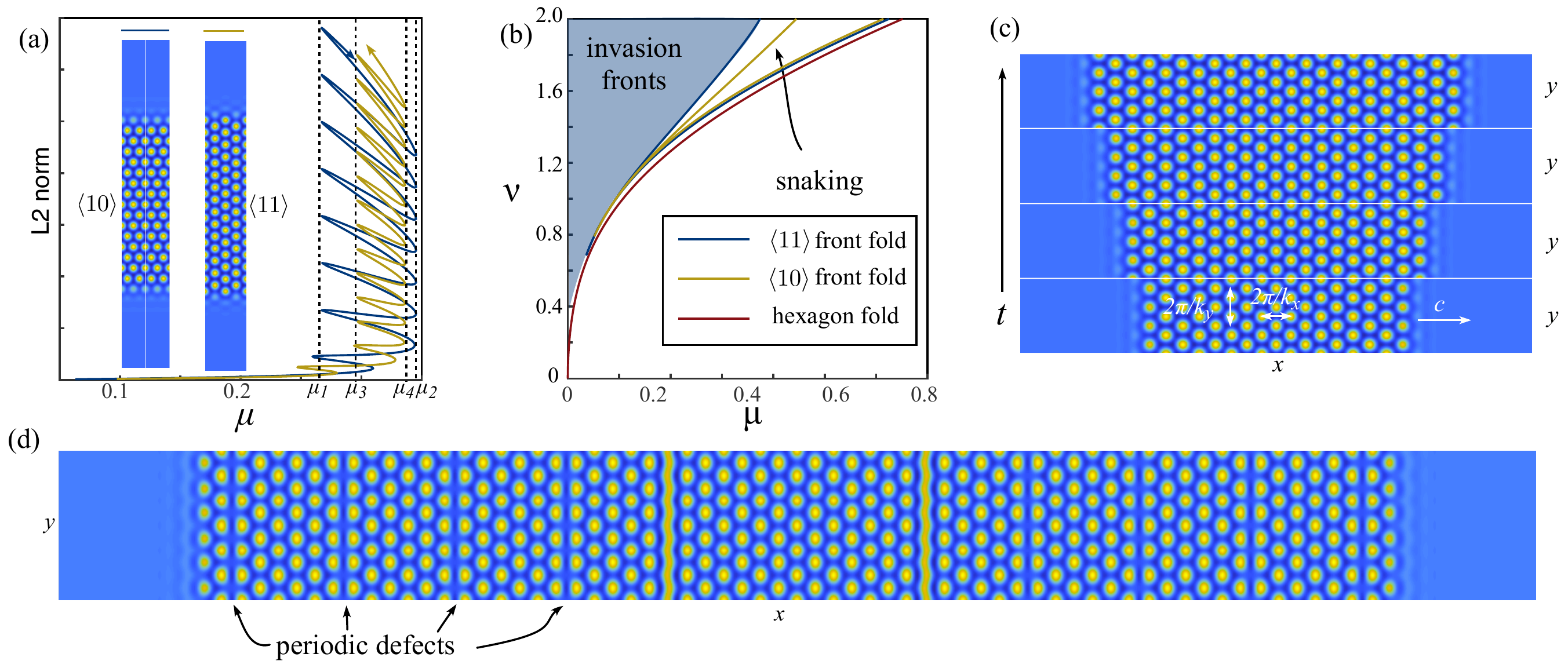}
\caption{(a) Snaking of the stationary hexagon $\langle10\rangle$- and $\langle 11\rangle$-fronts for $k_y=\frac12$ and $\sqrt{3}/2$. (b) $(\mu,\nu)$-bifurcation plot of the hexagon stationary front. (c) time simulation of a $\langle10\rangle$-hexagon front for $(\mu,\nu)=(0.05,1.6)$ with $k_y=0.86$ and (d) time simulation of a  $\langle10\rangle$-hexagon front with periodic defects for $(\mu,\nu)=(0.05,1.6)$ with $k_y=0.72$ at $t=300$ (the front has been periodically extended 3 times in $y$).\label{f:stat_hex_snake}}
\end{figure}

In 2008, Lloyd et al.\ \cite{lloyd2008} showed how stationary hexagon fronts undergo the process of homoclinic snaking; see Figure~\ref{f:stat_hex_snake}(a). Homoclinic snaking is where localized pulses involving a spatially periodic core develop successively wider periodic cores through infinitely-many folds, as a control parameter is varied. All the pulses co-exist in an open parameter space  where the trivial and domain covering cellular pattern states are bistable. While there are infinitely many possible hexagon fronts where the pattern is orientated with respect to the front interface, Lloyd et al.\ \cite{lloyd2008} concentrated on two main orientations, namely the $\langle10\rangle$- and the $\langle11\rangle$-fronts as defined by the Bravais-Miller index. It was found that these two types of fronts have different widths of snaking in parameter space; see Figure~\ref{f:stat_hex_snake}(a) where it is shown that the $\langle10\rangle$-fronts snake between $\mu_1\leq\mu\leq\mu_2$ while the $\langle11\rangle$-front snake between $\mu_3\leq\mu\leq\mu_4$ for $\nu=1.6$. The homoclinic snaking regions for the two fronts form wedges in the $(\mu,\nu)$-parameter space where to the left of the snaking region hexagons invade the trivial state and retreat on the right of the snake; see Figure~\ref{f:stat_hex_snake}(c) for an example of an invading hexagon $\langle10\rangle$-front. An explanation for the different widths of the hexagon fronts' snaking regions near $\nu\approx0$, was shown by Koyzreff \& Chapman using exponential asymptotics~\cite{kozyreff2013}. In particular, Koyzreff \& Chapman showed that for $\nu\approx0$ the hexagon fronts with the largest snaking regions are the $\langle10\rangle$- and the $\langle11\rangle$-fronts. 

\begin{figure}[h]
	\centering
		\includegraphics[width=\linewidth]{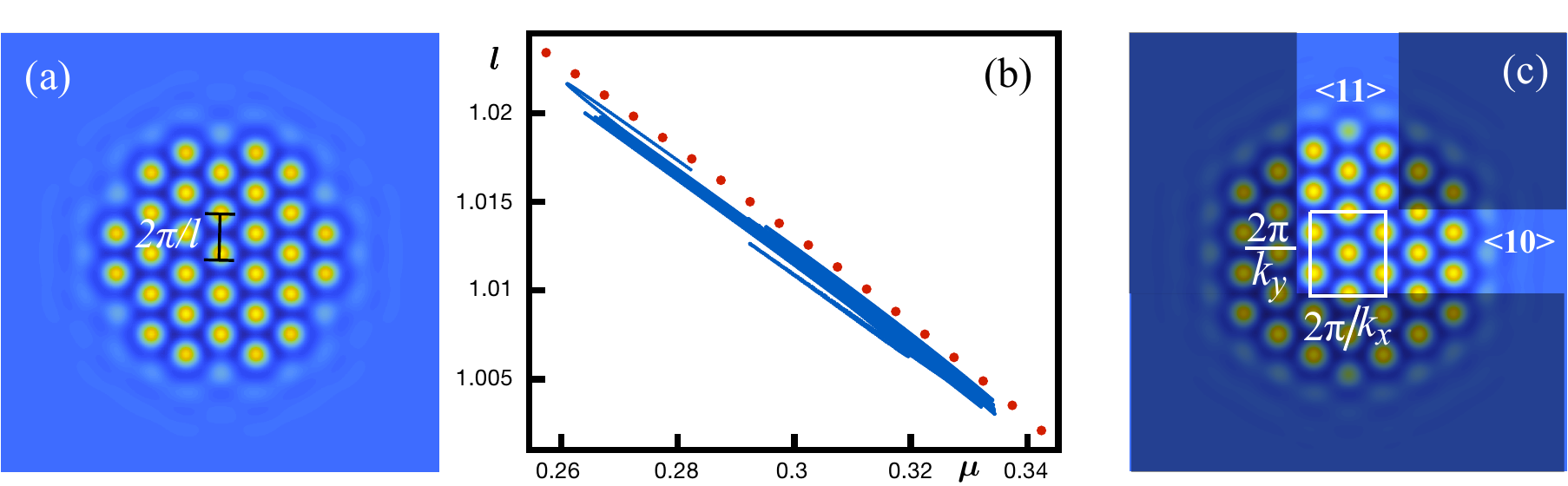}
	\caption{(a) Stationary patch wavenumber selection from time simulation numerics of the SH equation with $\nu=1.6$ compared with (b) the Hamiltonian wavenumber selected for the perfect $\mathbb{D}_6$ cellular hexagon. Figure reproduced from~\cite{lloyd2008}. Panel (c) depicts how a hexagon patch can be viewed as comprising of various $\langle10\rangle$- and $\langle11\rangle$-hexagon fronts orientated perpendicular to each other.\label{f:stat_patch_select}}
\end{figure}

Lloyd et al.\ \cite{lloyd2008} also provided a pattern selection criterion for the selected cellular far-field hexagons of a stationary planar hexagon front. The criterion provides a wavenumber, $k_x$, selection principle of the far-field hexagon wavenumber as a function of $k_y$ for a hexagon front connecting to the trivial state as shown in Figure~\ref{f:hex_front_patch}. However, it remains to explore how the selected wavenumber varied as a function of $k_y$; Lloyd et al.\ only investigated the cases $k_y=\frac12$ and $\frac{\sqrt{3}}{2}$ for their numerical explorations. 
Furthermore, they also showed from time simulations that fully localized hexagon patches select perfect $\mathbb{D}_6$-symmetric hexagons (see Figure~\ref{f:stat_patch_select}) as opposed to distorted hexagons but did not provide any possible explanation for this. 

Outside the snaking region with $\mu$ less than the homoclinic snaking region, it is found that the hexagon fronts and patches invade in a ``stick-slip'' or ``lurching'' manner where they develop a full set of hexagon cells along the interface before adding another set of cells. This process is similar to that observed in 1D~\cite{burke2007a,burke2007b,lloyd2019} and it is expected that the hexagon invading fronts select both a propagation speed, $c$, and far-field wavenumber, $k_x$, for a given $k_y$, while in the bistable region $\mu>0$; see Figure~\ref{f:stat_hex_snake} and Figure~\ref{f:stat_patch_select} and Lloyd~\cite{lloyd2019}. For $\mu$ greater than the homoclinic snaking region, the hexagon fronts and patches retreat, and it is expected that they select a propagation speed $c$. Interesting instabilities of invading hexagon fronts can also be found. In Figure~\ref{f:stat_hex_snake}(d), we see that stretching the hexagons in the transverse direction $y$  (equivalent to decreasing $k_y$) leads to a hexagon front that develops defects in the spatial pattern. We only observe this instability for the $\langle10\rangle$-front and not the $\langle11\rangle$-front. From a heuristic close-packing argument, the $\langle11\rangle$-front appears to be robust to stretching in the transverse direction as the horizontal rows of the hexagon cells are more closely packed than in the $\langle10\rangle$-front and so cannot be stretched in the horizontal direction as much. Due to the presence of these types of instabilities, it is natural to want to use numerical path-following techniques to explore this in a more systematic manner and understand what type of instability gives rise to this phenomenon. 

The aim of this paper is to investigate the pattern selection mechanism and any bifurcations of stationary and invading hexagon fronts in the bistable region. We also provide a heuristic criterion for the selected wavenumber of the cellular hexagons in the center of a fully localized patch (either invading or stationary); see Figure~\ref{f:stat_patch_select}. In Figure~\ref{f:stat_patch_select}(c) we propose that a hexagon patch can be viewed as being made up of $\langle10\rangle$- and $\langle11\rangle$-hexagon like fronts perpendicular to each other. Now given a $k_y=: k_y^{\langle10\rangle}$, the $\langle10\rangle$-front will select a $k_x^{\langle10\rangle}:=k_x(k_y^{\langle10\rangle})$ wavenumber for the cellular hexagon in the white box in Figure~\ref{f:stat_patch_select}(c). Similarly, the $\langle11\rangle$-front will select a $k_y^{\langle11\rangle}:=k_y(k_x^{\langle11\rangle})$ wavenumber for the hexagon in the white box for a prescribed $k_x=: k_x^{\langle11\rangle}$. For the cellular hexagon in the white box to be `compatible' with the two perpendicular fronts, we require $(k_x^{\langle10\rangle},k_y^{\langle10\rangle}) = (k_x^{\langle11\rangle}, k_y^{\langle11\rangle})$. It is expected that this condition to typically be met at isolated points in $(k_x,k_y)$-parameter space for the hexagon fronts.

We present our criterion as a conjecture. 

\begin{conjecture}\label{c:hex_conjecture}
Fix $(\mu,\nu)$ and $\mu>0$, then stationary and invading stable hexagon 2D patch solutions of (\ref{e:sh}) select a unique hexagon cellular pattern. The wavenumbers of the selected hexagon cellular pattern are determined by the compatibility of the selected far-field wavenumbers of the hexagon $\langle10\rangle$- $\langle11\rangle$-fronts orientated perpendicular to each other as shown in Figure~\ref{f:stat_patch_select}(c) and occurs when $(k_x^{\langle10\rangle},k_y^{\langle10\rangle}) = (k_x^{\langle11\rangle}, k_y^{\langle11\rangle})$. 
\end{conjecture}

We also verify the compatibility criterion for a non-variational SH equation. 
A key point here is that the numerical method presented in this paper allows one to compute the compatibility diagram in general reaction-diffusion systems without a variational/gradient structure. 

In a previous paper~\cite{lloyd2019}, we numerically investigated the invading fronts, whose far-field involved stationary stripes, using path-following routines. The numerical method used in~\cite{lloyd2019} employed a ``far-field core decomposition'' that has been recently developed in a series of papers~\cite{morrissey2015,lloyd2017,avery2019,weinburd2019,dodson2019}. Lloyd~\cite{lloyd2019} also showed that invading stripes should select a far-field wavenumber and a temporal invasion period. Adapting the arguments in~\cite{lloyd2019}, it is easy to show that hexagon invasion fronts in the bistable region to also select a far-field wavenumber $k_x$ for a fixed $k_y$. In this paper, we use the same numerical method to explore invading hexagon fronts and any bifurcations they may undergo. 

In the context of directional quenching, for a small spatial parameter jump i.e., $\mu = \epsilon\mbox{sign}(x-ct),\epsilon\ll1$ and $c$ is the speed of the quenching, hexagon fronts have been investigated by Weinburd in his PhD thesis~\cite{weinburd2019} using normal form analysis and numerical continuation for small $\epsilon$. There it is found for small speeds $c$, that the fronts select a wavenumber $k_x$ less than the linear critical one.  

The paper is outlined as follows. In \S\ref{s:hex_stab}, we review the existence and stability theory for weakly nonlinear cellular distorted hexagons and distorted hexagons far away from onset. Section~\ref{s:pat_select_stat} reviews the selection principle for hexagon fronts, investigates the effect of varying $k_y$ of the hexagon fronts on the homoclinic snaking region and introduces the concept of a compatibility diagram. In \S~\ref{s:weak} we review the weakly nonlinear theory for hexagon fronts in the SH equation. Section~\ref{s:method} we describe the numerical algorithms used to compute hexagon invasion fronts, and we present our results on the pattern selection of these fronts in \S\ref{s:numerics}. We then compare our results and predictions from time simulations of an invading hexagon patch in the SH equation in \S\ref{s:patch_invasion} and finally conclude in \S\ref{s:discussion}.

\section{Hexagon Existence and Stability}\label{s:hex_stab}
In this section, we review the existence of small amplitude (distorted) hexagons and their stability in the planar SH equation for $\nu\approx0$; see~\cite{gunaratne1994,pena2001}. 

Regular cellular hexagons lie on the planar hexagonal lattice $\mathcal{L}$
\[
\mathcal{L} = \left\{n_1l_1+n_2l_2+n_3l_3\in\mathbb{R}^2\;:\:  n_1,n_2,n_3\in\mathbb{Z}\right\},
\]
where, 
\[
l_1 = 2\pi\kappa\left(1,\frac{1}{\sqrt{3}} \right),\qquad l_2 = 2\pi\kappa\left(0,-\frac{2}{\sqrt{3}} \right),\qquad l_3 = 2\pi\kappa\left(-1,\frac{1}{\sqrt{3}} \right),
\]
and $\kappa$ is the wavenumber. The dual lattice $\mathcal{L}^*$ is given by
\[
\mathcal{L}^* = \left\{n_1k_1+n_2k_2+n_3k_3\in\mathbb{R}^2\;:\:  n_1,n_2,n_3\in\mathbb{Z}\right\},
\]
where, 
\[
k_1 = \kappa\left(-1,0\right),\qquad k_2 =\kappa\left(\frac12,\frac{\sqrt{3}}{2}\right),\qquad k_3 =\kappa\left(\frac12,-\frac{\sqrt{3}}{2}\right),
\]
and we may represent all $\mathcal{L}$-periodic solutions of~(\ref{e:sh}) by the Fourier series,
\[
u(\mathbf{x},t) = \sum_{k\in\mathcal{L}^*}\hat u_k(t)e^{ik\cdot\mathbf{x}}.
\]

\begin{figure}[h]
	\centering
	\includegraphics[width=0.8\linewidth]{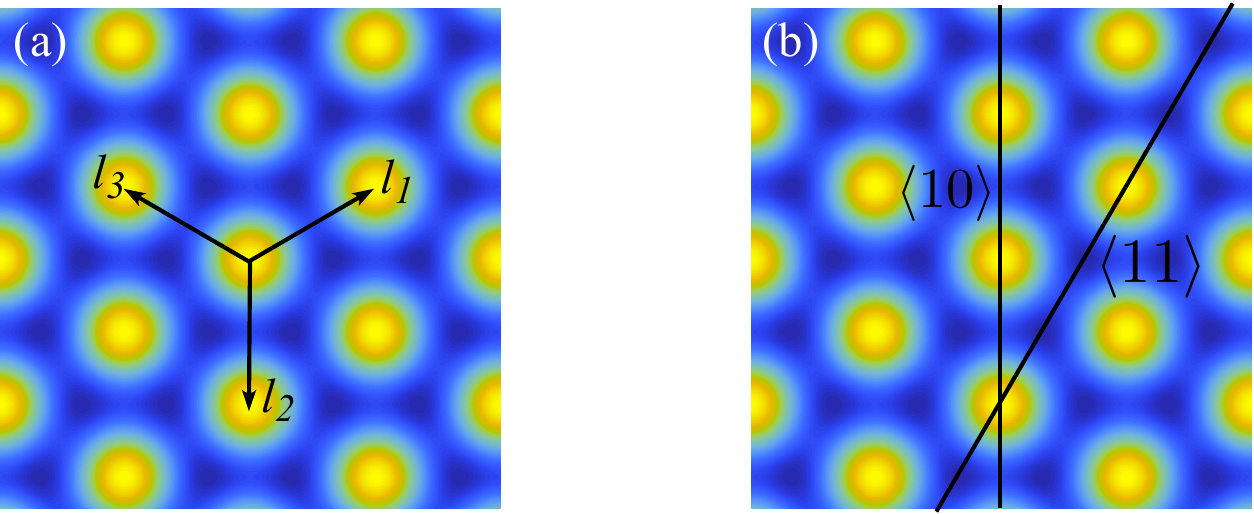}
	\caption{(a) shows the hexagon lattice vectors $l_1,l_2$ and $l_3$ on the cellular hexagon pattern and (b) shows the two lines with Bravais-Miller indices $\langle10\rangle$ and $\langle11\rangle$.\label{f:bravaismiller}}
\end{figure}

Following the convention in~\cite{lloyd2008}, define two types of possible interfaces (either $\langle10\rangle$- or $\langle11\rangle$-interfaces) on a fixed hexagonal lattice using the Bravais-Miller index~\cite{ashcroft1976}; see Figure~\ref{f:bravaismiller}. For distorted hexagons we keep this naming convention. 

To prove the existence of distorted hexagons bifurcating from $u=0$, we first rescale $(x,y)$ by setting $x = X/\kappa_1$ and $y = Y/\kappa_2$. In the rescaled variables the stationary quadratic-cubic SH equation becomes
\[
0 = -(1 + \kappa_1^2\partial_X^2 + \kappa_2^2\partial_Y^2)^2u - \mu u + \nu u^2 - u^3.
\] 
We consider the nonlinear problem $F(\mu,\nu,\kappa_1,\kappa_2,u) = 0$, where
\begin{equation}\label{e:non_hex}
F(\mu,\nu,\kappa_1,\kappa_2,u) := -(1 + \kappa_1^2\partial_X^2 + \kappa_2^2\partial_Y^2)^2u - \mu u + \nu u^2 - u^3, \qquad F:\mathbb{R}^4\times \mathcal{X}\rightarrow \mathcal{Y},
\end{equation}
and,
\begin{align*}
\mathcal{X} =& \left\{u\in H^4_{\mbox{loc}}(\mathbb{R}^2,\mathbb{R});\; u(X,Y) = u\left(X,Y+\frac{4\pi}{\sqrt{3}}\right) = u\left(X+2\pi,Y + \frac{2\pi}{\sqrt{3}}\right),\; \forall (X,Y)\right\},\\
\mathcal{Y} =&  \left\{u\in L^2_{\mbox{loc}}(\mathbb{R}^2,\mathbb{R});\; u(X,Y) = u\left(X,Y+\frac{4\pi}{\sqrt{3}}\right) = u\left(X+2\pi,Y + \frac{2\pi}{\sqrt{3}}\right),\; \forall (X,Y)\right\}.
\end{align*}

The linearization about $u=0$ is given by $Lu = \partial_uF(0,0,1,1,0)[u] = -(1 + \partial_X^2 + \partial_Y^2)^2u $ is self-adjoint with a discrete spectrum and kernel $\mbox{ker}(L)$ spanned by 6 eigenfunctions given by $u_1=e^{iX}$, $u_2=e^{-\frac12X + \frac{\sqrt{3}}{2}Y}, u_3 = e^{-\frac12X - \frac{\sqrt{3}}{2}Y}$ and their complex conjugates. We use the $L^2$-inner product on the fundamental periodicity domain $[0,4\pi]\times[0,4\pi/\sqrt{3}]$ to define the complements of $\mathcal{X}$ and $\mathcal{Y}$. In particular, we assume the decompositions 
\[
\mathcal{X} = \mbox{ker}(L) \oplus \mbox{ker}(L)^\perp,\qquad \mbox{and}\qquad \mathcal{Y} = \mbox{ran}(L) \oplus  \mbox{ran}(L)^\perp,
\]
and define the orthogonal projection $Pu = \frac{1}{8\pi^2\sqrt{3}}\sum_{i=1}^6\langle u_i,u\rangle u_i$, where $\langle u,v\rangle = \int_0^{4\pi/\sqrt{3}}\int_0^{4\pi}u\bar v\D x\D y$ and the mapping $Q\; \colon \: \mathcal{X} \mapsto \mathbb{C}^6$ where $Qu = \frac{1}{8\pi^2\sqrt{3}}\left(\langle u,u_1\rangle,\ldots, \langle u,u_6\rangle \right)^T$.

We decompose $u\in\mathcal{X}$ into $(\sum_{i=1}^3\alpha_i u_i + c.c.) + V$, where $\alpha_i\in\mathbb{C}$, and $PV =0 $. We re-write the nonlinear problem as
\begin{align}
QF(\mu,\nu,\kappa_1,\kappa_2,(\sum_{i=1}^3\alpha_i u_i + c.c.) + V) =& \mathbf{0},\\
(I-P)F(\mu,\nu,\kappa_1,\kappa_2,(\sum_{i=1}^3\alpha_i u_i  + c.c.) + V) =& 0.
\end{align}
The second equation can be solved for $V\in(I-P)\mathcal{X}$ as a function of $(\mu,\nu,\kappa_1,\kappa_2,\alpha)$ using the implicit function theorem. From $(I-P)Lu_j = 0$, we find that $V=G(\nu,\kappa_1,\kappa_2,\alpha) = \mathcal{O}(\nu|\alpha|^2 + |\alpha|^3)$ for $\alpha\rightarrow 0$. Hence, all small solutions of (\ref{e:non_hex}) satisfy the bifurcation equations
\[
QF(\mu,\nu,\kappa_1,\kappa_2,(\sum_{i=1}^3\alpha_i u_i + c.c.) + V) = \mathbf{0},
\]
given by
\begin{subequations}\label{e:hex_normalform}
\begin{align}
\Gamma_1\alpha_1 + \mu\alpha_1 + \beta_1\overline{\alpha_2\alpha_3} - \beta_2|\alpha_1|^2\alpha_1 -\beta_3(|\alpha_2|^2+|\alpha_3|^2)\alpha_1+\mathcal{O}(|\alpha|^4) =& 0,\\
\Gamma_2\alpha_2 + \mu\alpha_2 + \beta_1\overline{\alpha_1\alpha_3} - \beta_2|\alpha_2|^2\alpha_2 -\beta_3(|\alpha_1|^2+|\alpha_3|^2)\alpha_2+\mathcal{O}(|\alpha|^4) =& 0,\\
\Gamma_3\alpha_3 + \mu\alpha_3 + \beta_1\overline{\alpha_2\alpha_1} - \beta_2|\alpha_3|^2\alpha_3 -\beta_3(|\alpha_2|^2+|\alpha_1|^2)\alpha_3+\mathcal{O}(|\alpha|^4) =& 0,
\end{align}
\end{subequations}
and the complex conjugates of the equations, 
where $\Gamma_1 = -(\kappa_1^2-1)^2,\Gamma_2=-\frac{1}{16}(\kappa_1^2+3\kappa_2^2-4)^2,\Gamma_3=\Gamma_2$, and
\[
\beta_1 = 2\nu+\mathcal{O}(|\mu|(|\mu| + |\nu|)),\qquad \beta_2 = 3+\mathcal{O}(|\mu|+|\nu|),\qquad \beta_3 = 6 + \mathcal{O}(|\mu|+|\nu|).
\] 
The equations~(\ref{e:hex_normalform}) have been analyzed formally in~\cite{malomed1994,matthews1998,pena2001} and we state some of the key results. 

We are interested in rectangular perturbations in $x$ and $y$. Hence, we set $(\kappa_1,\kappa_2)=(\sqrt{1+\delta_x},\sqrt{1+\delta_y})$ and $\alpha_2=\alpha_3$. Furthermore, we introduce the following scalings
\[
\mu = \epsilon^2\tilde\mu,\qquad\nu=\epsilon\tilde\nu,\qquad\delta_x=\epsilon\tilde\delta_x,\qquad\delta_y = \epsilon\tilde\delta_y,\qquad\alpha_1 = \epsilon A,\qquad \alpha_2 = \epsilon B.
\]
Hence, 
real equilibria satisfy the equations
\begin{subequations}\label{e:hex_amp_eq}
\begin{align}
\mu_1 A +2\tilde\nu B^2 - 3A^3 - 12B^2A + \mathcal{O}(\epsilon)=& 0,\\
\mu_2 +2\tilde\nu A - 9B^2 - 6A^2 + \mathcal{O}(\epsilon) =&0,
\end{align}
\end{subequations}
where $(\tilde\mu_1,\tilde\mu_2) = (\tilde\mu-\tilde\delta_x^2,\tilde\mu - \frac{(\tilde\delta_x+3\tilde\delta_y)^2}{16})$. Solving for $B^2$ in the second equation and substituting into the first, yields a cubic equation for $A$
\[
a_3A^3 + a_2A^2 + a_1A + a_0 = 0,
\]
where $a_0=2\tilde\nu\tilde\mu_2+ \mathcal{O}(\epsilon), a_1=(4\tilde\nu^2+9\tilde\mu_1-12\tilde\mu_2)+\mathcal{O}(\epsilon), a_2=-36\tilde\nu+\mathcal{O}(\epsilon),a_3=45+\mathcal{O}(\epsilon)$. Setting $\epsilon=0$, the discriminant of the cubic $D(\tilde\mu_1,\tilde\mu_2,\tilde\nu,\epsilon)=0$ determines the boundary between the cubic have one ($D(\tilde\mu_1,\tilde\mu_2,\tilde\nu,\epsilon)<0$) or three real roots ($D(\tilde\mu_1,\tilde\mu_2,\tilde\nu,\epsilon)>0$). For $B^2=0$, there are two cases; (a) $A=0$ and $\mu_2=0$ corresponding to the bifurcation from the trivial state and (b) $A=\pm\sqrt{\tilde\mu_1/3}$ and $\tilde\mu_2=2\tilde\mu_1\mp2\tilde\nu\sqrt{\tilde\mu_1/3}$ corresponding to the bifurcation from the stripe states. 

\begin{Proposition}\label{p:p1}
There exists an $\epsilon_0>0$ such that for all $\epsilon\in(0,\epsilon_0]$ and $D(\tilde\mu_1,\tilde\mu_2,\tilde\nu,\epsilon)>0$ where
\[
D(\tilde\mu_1,\tilde\mu_2,\tilde\nu,\epsilon) = -27a_0^2a_3^2+(18a_0a_1a_2-4a_1^3)a_3-4a_0a_2^3+a_1^2a_2^2
\]
there exists a rectangular distorted hexagon solution to equation (\ref{e:sh}).
\end{Proposition}

\begin{figure}[h]
	\centering
	\includegraphics[width=0.8\linewidth]{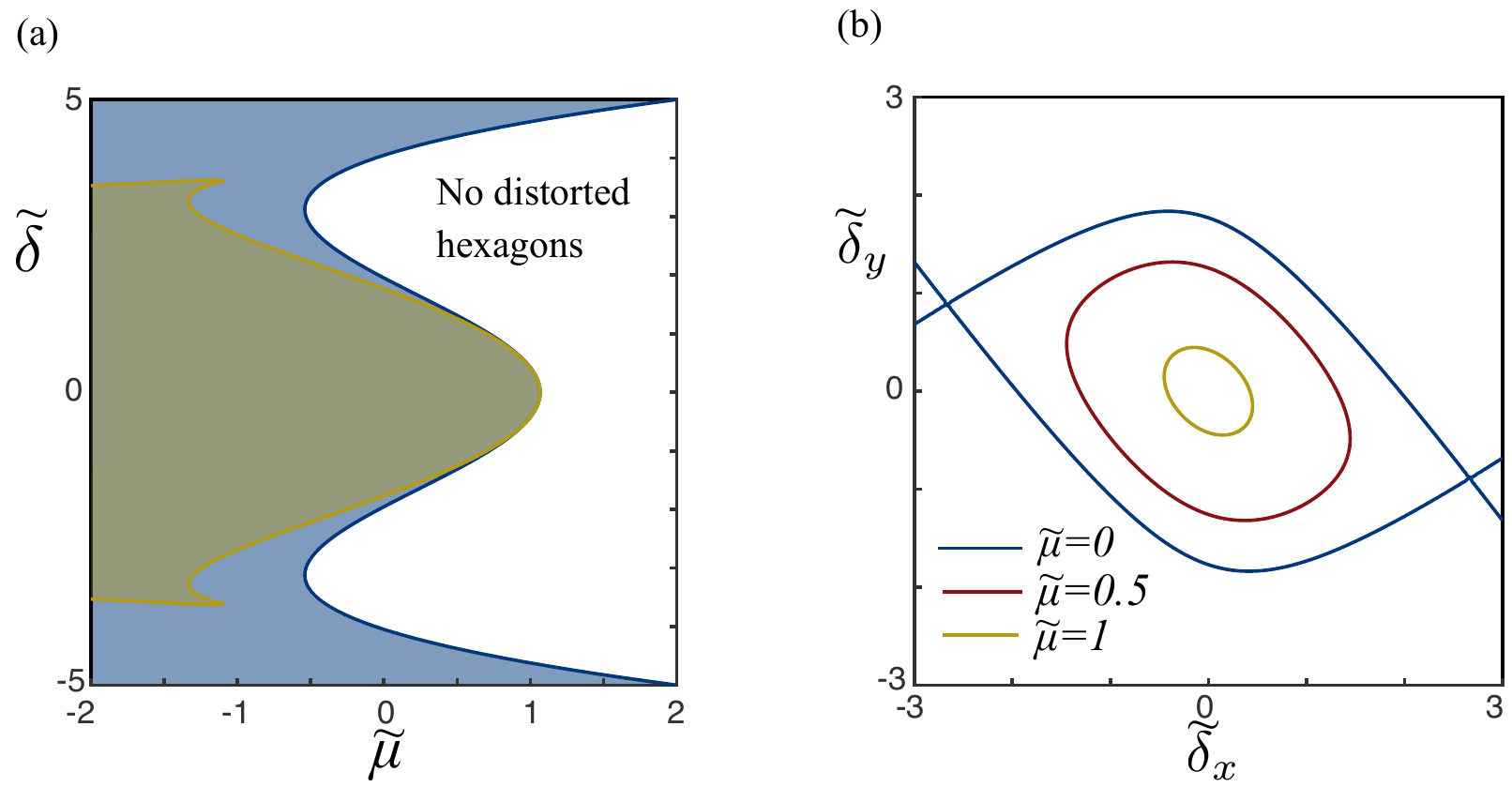}
	\caption{(a) Existence boundaries of the distorted hexagons from the amplitude equation analysis for $\tilde\nu=4$ as defined in Proposition~\ref{p:p1} with $\epsilon=0$. The blue shading region depicts the existence region of the distorted hexagon due to perturbations in the $x$-direction i.e., $\tilde\delta_y=0$ and the gold shaded region is the existence region in the $y$-direction i.e., $\tilde\delta_x=0$. (b) Plot of the existence boundaries of the distorted hexagons from the amplitude equation analysis for different $\tilde\mu$ level sets in $(\tilde\delta_x,\tilde\delta_y)$-space with $\tilde\nu=4$. Existence of distorted hexagons lies in the bounded regions. We see that we can slightly perturb a hexagon more in the $x$-direction than in the $y$-direction.\label{f:hex_distort}}
\end{figure}
In Figure \ref{f:hex_distort}(a), we plot the existence boundaries of the distorted hexagons from the discriminant $D(\tilde\mu_1,\tilde\mu_2,\tilde\nu,0)$ with solely $x$- or $y$-perturbations for $\tilde\nu=4$. We see that the regular hexagons are the last to undergo a fold as $\tilde\mu$ is increased in the bistable region. The $y$-distorted hexagons (corresponding to $\langle11\rangle$-perturbations) shaded in gold exist for a smaller region in parameter space than the $x$-perturbed hexagons (corresponding to $\langle 11\rangle$-perturbations). In Figure \ref{f:hex_distort}(b), we plot the existence boundaries in $(\tilde\delta_x,\tilde\delta_y)$-space for different values of $\tilde\mu$. Here we see that the most extreme form of distorted hexagons occurs when $\tilde\delta_x$ and $\tilde\delta_y$ are of opposite signs corresponding to squeezed hexagons in one direction and stretched in the other. 

\begin{figure}[h]
	\centering
	\includegraphics[width=\linewidth]{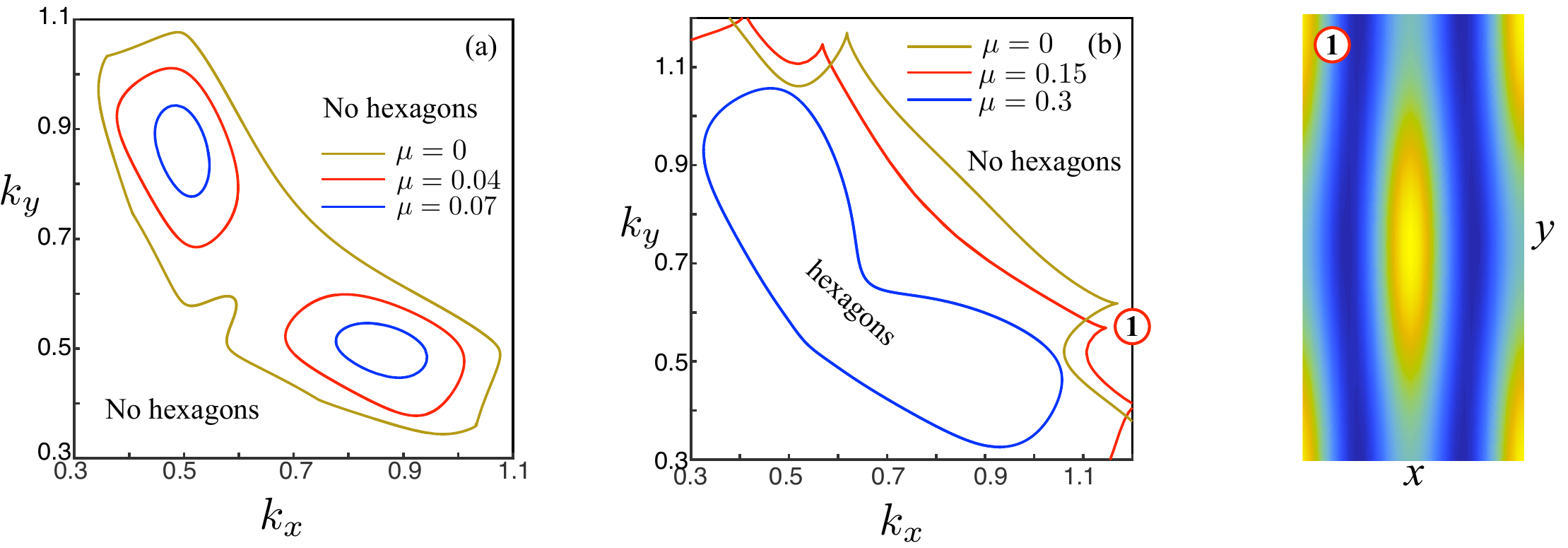}
	\caption{(a) Existence boundaries for the SH equation with $\nu=0.9$ and $\mu=0,0.04,0.07$. (b) Existence and stability boundaries for distorted hexagons in the quadratic-cubic SH equation with $\nu=1.6$ and $\mu=0,0.15,0.3$. Panel {\rm \raisebox{.5pt}{\textcircled{\raisebox{-.9pt} {1}}}} shows a plot of the distorted hexagon at the cusp shown in (b).\label{f:hex_floquet}}
\end{figure}
Further away from $\nu\approx0$, we numerically compute the distorted hexagons for $\nu=0.9$ and $1.6$ for various $\mu$ values in Figure~\ref{f:hex_floquet}(a) and \ref{f:hex_floquet}(b). The computations are done on a doubly periodic square box using a Fourier pseudo-spectral method (see~\cite{lloyd2008}) such that $u(k_xx,k_yy)=u(\xi,\eta)$ where $\xi,\eta\in(0,2\pi]$. For $\nu=0.9$, we see for large $\mu$ values that the distorted hexagons exist in closed regions in $(k_x,k_y)$-space around the linear critical wavenumbers similar to that predicted from the asymptotics in Figure~\ref{f:hex_distort}(b). For $(\mu,\nu)=(0,0.9)$, the two existence boundaries merge into one and now crosses the $k_x=k_y$ line corresponding to squares. For $\nu=1.6$, we see for $\mu=0.3$ (in the middle of the snaking regions for both the $\langle10\rangle$- and $\langle11\rangle$-fronts) the existence boundary of the distorted hexagons is a smooth curve and includes the square case. For $\mu=0$ and $0.15$ in Figure~\ref{f:hex_floquet}(b), the existence boundary becomes highly complex with cusps in the boundary, and we just plot the boundaries in the region $(k_x,k_y)\in[0.3,1.2]^2$ (though one can find distorted hexagons outside this region). In panel \raisebox{.5pt}{\textcircled{\raisebox{-.9pt} {1}}}, we plot the hexagon solution at one of the cusps in original $(x,y)$ coordinates showing that the cusp is occurring where the hexagons bifurcate off the stripes. 

\begin{figure}[h]
	\centering
	\includegraphics[width=\linewidth]{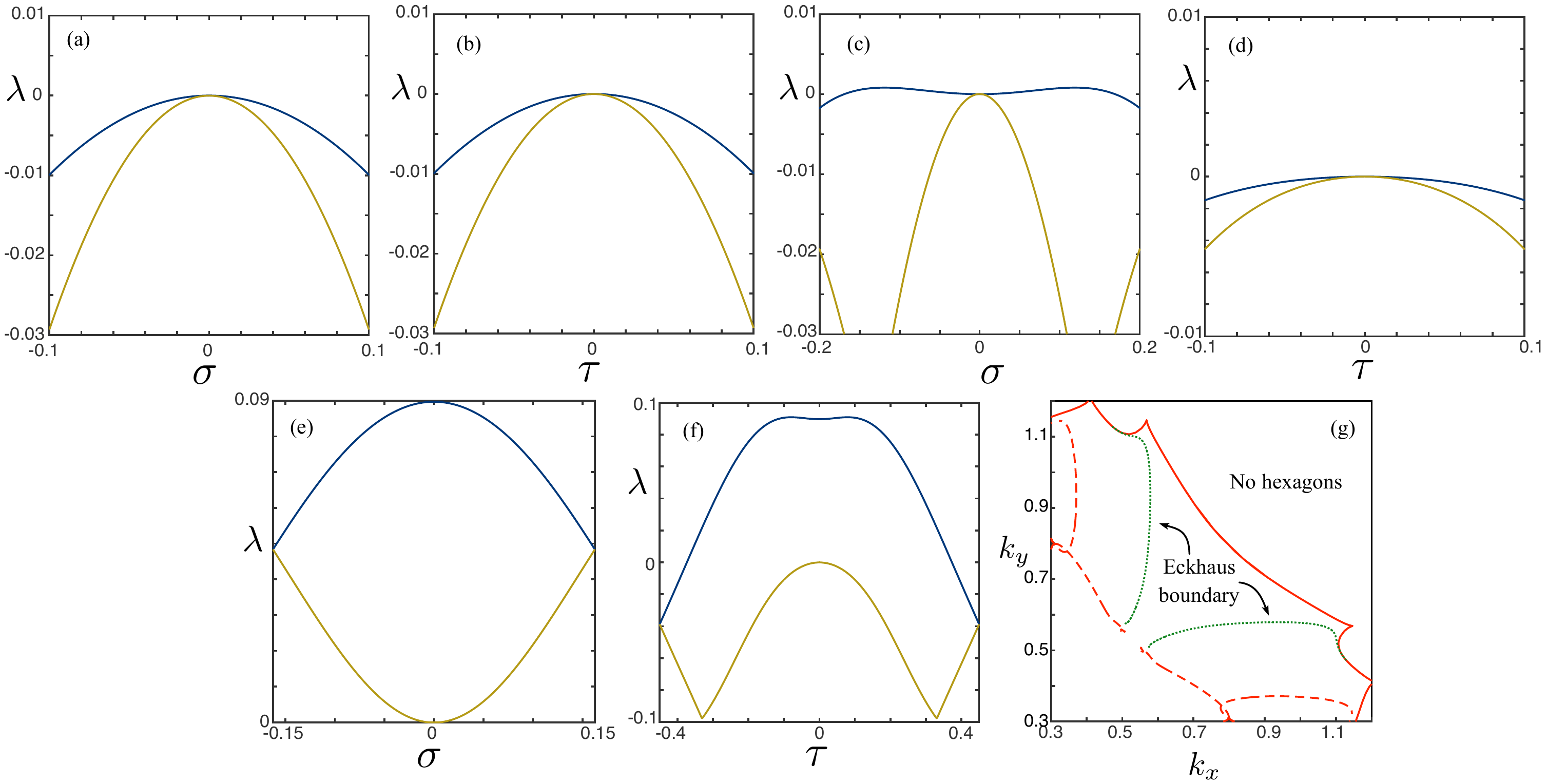}
	\caption{Plots of the two largest Bloch eigenvalues that remain non-positive of the (a) $\langle10\rangle$-hexagon and (b) $\langle11\rangle$-hexagon computed on the rescaled periodic domain $[0,2\pi)^2$ with $(\mu,\nu)=(0.1,1.6)$ and $(k_x,k_y)=(\frac12,\frac{\sqrt{3}}{2})$. All other eigenvalues are strictly negative for all $\sigma$ and $\tau$. Panels (c) and $(d)$ show plots of the two largest Bloch eigenvalues for a $\langle10\rangle$ and $\langle11\rangle$-hexagon (respectively) with $(k_x,k_y)=(0.5,0.675)$ and $(\mu,\nu)=(0.15,1.6)$. Panels (e) and (f) show plots of the two largest eigenvalues for a $\langle10\rangle$ and $\langle11\rangle$-hexagon (respectively) with $(k_x,k_y)=(0.325,0.9)$ and $(\mu,\nu)=(0.15,1.6)$. Panel (g) is a plot of the existence and some of the stability boundaries of the distorted hexagons where co-periodic boundaries are shown as dashed red lines and the Eckhaus boundaries are shown as dashed green lines. The computations are done using a Fourier pseudo-spectral method with 30 collocation points in $x$ and $y$ on the rescaled square periodic domain $(0,2\pi]^2$.\label{f:hex_floquetb}}
\end{figure}
Linear stability of the distorted hexagons with respect to rectangular perturbations can be found by looking at the linear problem
\[
-(1+\kappa_x^2\partial_X^2+\kappa_2^2\partial_Y^2)^2w - \mu w + (2\nu \tilde u - 3\tilde u^2)w = \lambda w.
\]
We restrict to perturbations of the form $w=e^{i(\sigma x + \tau y)}\tilde w(X,Y)$, $\tilde w\in \mathcal{X}$ leading to the Bloch eigenvalue problem
\[
L\tilde w :=-(1+(\kappa_x\partial_X+i\sigma)^2+(\kappa_2\partial_Y+i\tau)^2)^2w - \mu w + (2\nu \tilde u - 3\tilde u^2)w - \lambda w= 0,
\]
where $\sigma,\tau\in\mathbb{R}$. 
We have found that the stability of distorted hexagons can be highly complicated and instead of giving a thorough investigation we instead list some basic observations about the linear stability for distorted hexagons. For $\sigma=\tau=0$, the zero eigenvalue has algebraic multiplicity 2 with corresponding eigenfunctions given by $\tilde u_x$ and $\tilde u_y$. For regular and slightly distorted hexagons, the first two Bloch eigenvalues are quadratic in $\sigma$ and $\tau$ and typically look like those shown in Figure \ref{f:hex_floquetb}(a) and (b). The linear co-periodic stability i.e., $\sigma=\tau=0$ for the SH equation was computed by~\cite{matthews1998} for large amplitude hexagons and in the bistable region the stability and existence regions are typically found to be very close together. Phase instabilities of distorted hexagons have been studied in~\cite{pena2001} when the hexagons are distorted along the $y$ direction. It has been shown that the distorted hexagons for $\nu\approx0$ are stable with respect to spatially homogeneous perturbations in the bistable region; see for instance~\cite{pena2001}. We are able to numerically find the equivalent of an Eckhaus instability for distorted hexagons where an eigenvalue $\lambda$ has the expansion
\[
\lambda = c_\sigma \sigma^2 - c_\tau\tau^2 + \mathcal{O}((\sigma+\tau)^4),
\]
with $c_\sigma,c_\tau>0$; see for example Figure~\ref{f:hex_floquetb}(c). In Figure~\ref{f:hex_floquet}(b), we plot the existence and stabilities boundaries in $(k_x,k_y)$-space for $\mu=0.15$ and $\nu=1.6$. The stability computed with respect to co-periodic perturbations is shown as a dashed red line while the Eckhaus stability boundary is shown as a dashed green line. We find for large values of $\mu$ close to the fold of the hexagons (for example $\mu=0.3$) that the stability boundary is close to the existence curve. For smaller values of $\mu$, the existence and stability curves become larger and more complicated. There are many eigenvalues that can become unstable as the hexagons become strongly distorted.

\section{Pattern Selection Principle for Stationary Hexagon Fronts}\label{s:pat_select_stat}
In this section, we review the pattern selection criterion of Lloyd et al.\ \cite{lloyd2008} for stationary hexagon fronts and use it to explore the dependence of the hexagon fronts snaking region on $k_y$. Furthermore, we introduce the idea of a compatibility diagram for hexagon patches. 

We view a front as a heteroclinic orbit in $x$ that connects a cellular (possibly distorted) hexagon pattern $u_h(x,y;k_x,k_y)$  as $x\rightarrow-\infty$ to the trivial state as $x\rightarrow+\infty$ and is periodic in $y$; see Figure~\ref{f:hex_front_patch}. 

\begin{figure}[h]
	\centering
	\includegraphics[width=0.8\linewidth]{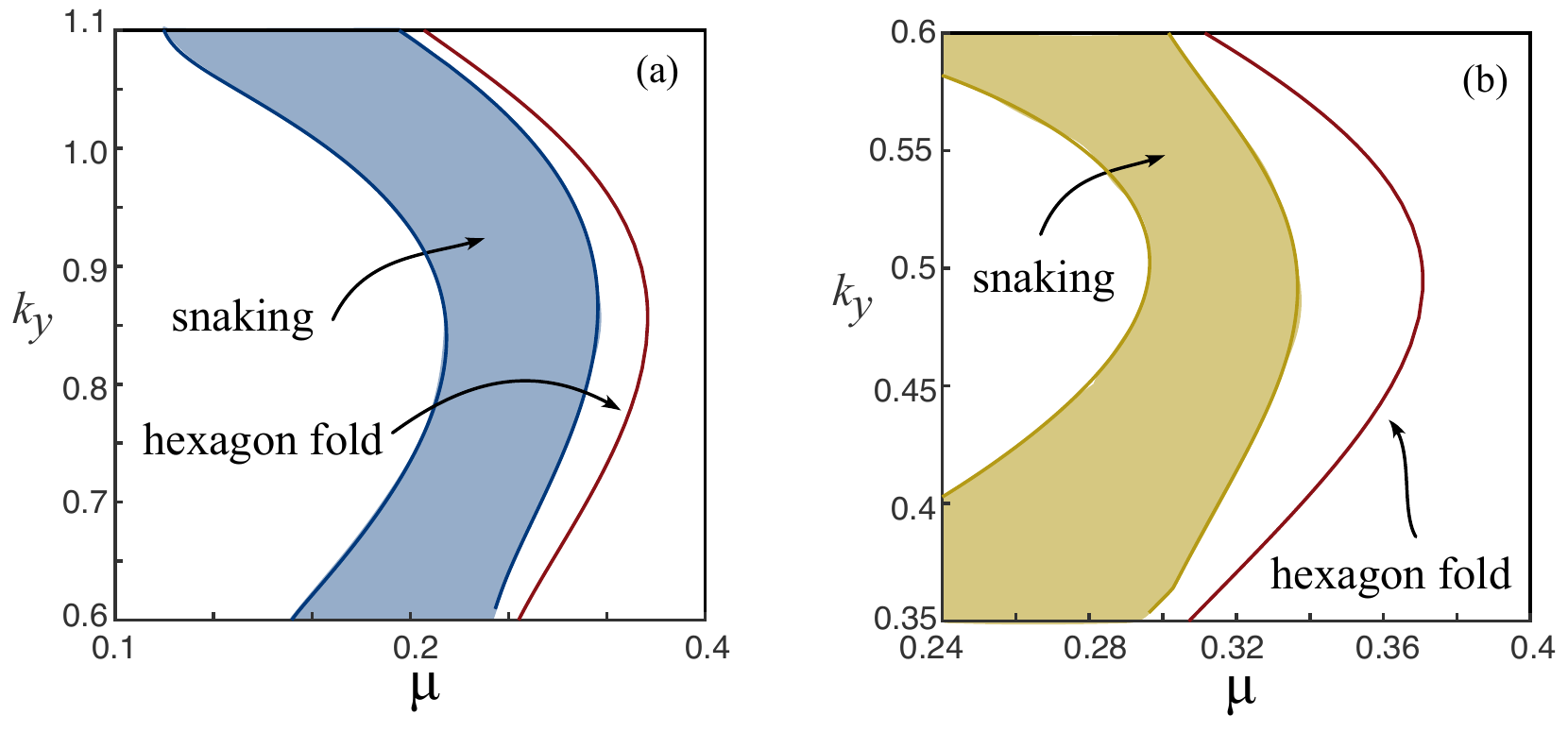}
	\caption{Widths of the snaking regions depicted as shaded regions for (a) $\langle10\rangle-$ and (b) $\langle 11\rangle-$ hexagon fronts with $\nu=1.6$. The fold of the cellular hexagon patterns is shown as a red line.\label{f:hex_snake_ky}}
\end{figure}
We start by exploring the dependence of the hexagon front snaking regions on $k_y$ for $\nu=1.6$. We note that this was not carried out by Lloyd et al.\ \cite{lloyd2008}. In Figure~\ref{f:hex_snake_ky}, we plot the snaking regions for the $\langle10\rangle$- and $\langle11\rangle$-fronts where $u(x,k_yy) = u(x,\xi)$ and $\xi$ is $2\pi$-periodic. The snaking regions are defined by the $\mu$-values of the location of the folds of the stationary fronts. We trace out the folds by taking a fold of a wide front (far up the homoclinic snake) and continuing it in $(\mu,k_y)$-parameter space. The snaking region extends the furthest into positive $\mu$ for $k_y$ around the linear critical values of $k_y=\frac{\sqrt{3}}{2}$ and $\frac12$ for the $\langle10\rangle$- and $\langle11\rangle$-fronts, respectively. It is clear that the widths of the snaking regions in $\mu$-parameter space dependent  heavily on $k_y$ and that the $\langle10\rangle$-fronts exist over a larger $k_y$ range.

Lloyd~et al.\ \cite{lloyd2008} also derived a selection criterion for the wavenumber of the hexagon in a localized pattern for a fixed $k_y$. We define the spatially ($x$ independent) conserved quantity $\mathcal{H}$,
\begin{equation}\label{e:ham}
\mathcal{H}(u) = \int_0^{L}\left[u_{xxx}u_x -\frac{u_{xx}^2}{2} + u_x^2 + \frac{(1+\mu)u^2}{2} - \frac{\nu u^3}{3} + \frac{u^4}{4} - u_{xy}^2-u_y^2+\frac{u_{yy}^2}{2}\right]\D y
\end{equation}
where $u(x,y)$ is a smooth solution of the Swift-Hohenberg equation~(\ref{e:sh}) which is spatially periodic with period $L$ in the $y$-variable; see~\cite{lloyd2008}. We call $\mathcal{H}$ the Hamiltonian analogous to the 1D stationary SH equation Hamiltonian. Since the conserved quantity $\mathcal{H}$ is independent of $x$, then for a planar distorted hexagon front that connects cellular hexagons to the trivial state, $\mathcal{H}$ must vanish when evaluated along a single hexagon, $u_h(x,y;k_x,k_y)$, in the far-field of the front.  For a hexagon front that is periodic in $y$ with period $2\pi/k_y$, then the front selects a unique wavenumber $k_x$ such that, 
\begin{equation}\label{e:stat_hex_ham}
\mathcal{H}(u_h(x,y;k_x,k_y)) = 0.
\end{equation}

\begin{figure}[h]
	\centering
	\includegraphics[width=\linewidth]{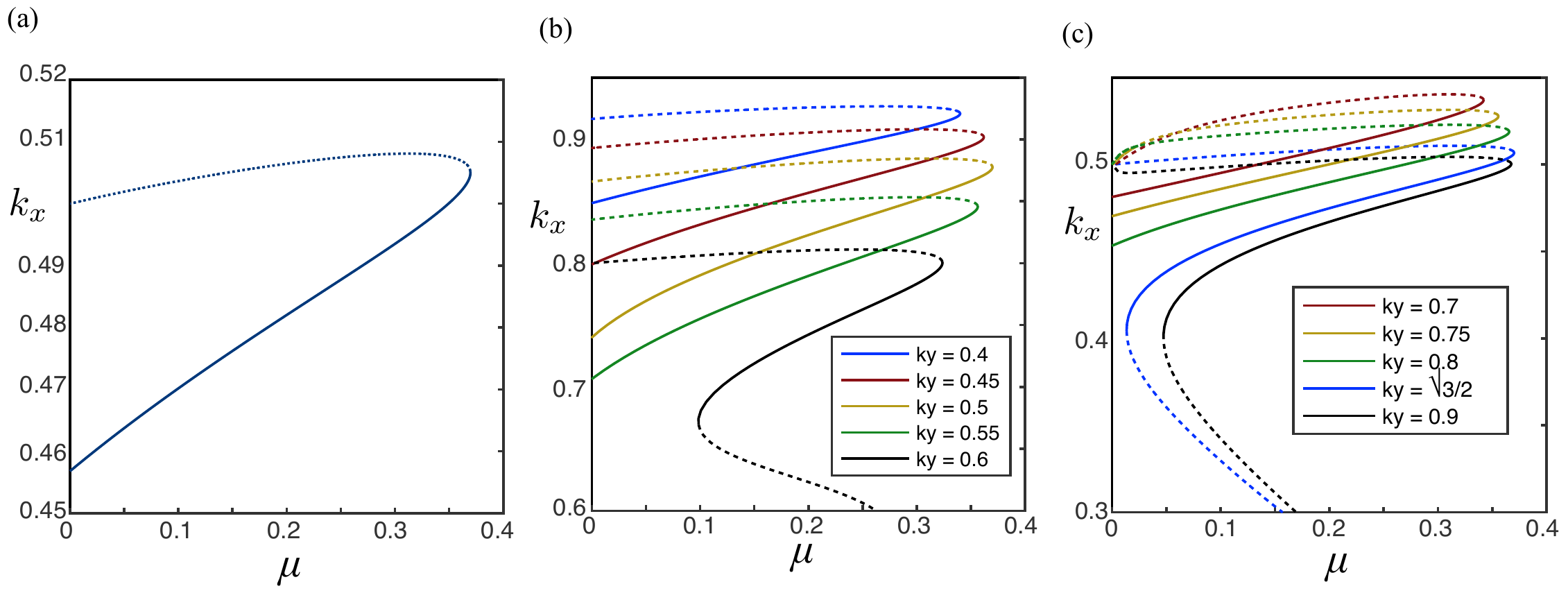}
	\caption{(a) Perfect Hexagon $\nu=1.6$ Hamiltonian wavenumber selection. $k_y=\sqrt{3}k_x$. (b) Distorted $\langle10\rangle$-hexagon and (c) $\langle11\rangle$-hexagon with  $\nu=1.6$. Co-periodic stability is shown as a solid line and co-periodic instability shown as dotted lines.\label{f:hex_ham}}
\end{figure}
In Figure~\ref{f:hex_ham}, we plot the selected wavenumbers for a perfect hexagon and distorted hexagons using (\ref{e:ham}) for $\nu=1.6$. At $(\mu,k_x)=(0,0.5)$, the hexagon cell bifurcates off the trivial state and is initially unstable. The selected wavenumber initially increases before going around a fold where the hexagon cell restabilizes with respect to co-periodic perturbations and the selected wavenumber decreases. The existence of a perfect hexagon i.e., $k_y=\sqrt{3}k_x$, satisfying the Hamiltonian constraint $\mathcal{H}$ for $\mu,\nu\sim0$ was proved in~\cite{lloyd2008}. In Figure~\ref{f:hex_ham}(a), we see that the selected wavenumber for a perfect hexagon behaves much the same way as for 1D stripes; see Burke and Knobloch~\cite{burke2007a}. When we allow the hexagons to be distorted in one direction (as shown in Figure~\ref{f:hex_ham}(b) and (c)), then we see that the primary fold location decreases in $\mu$ and for sufficiently large distortions, secondary folds occur near $\mu=0$. 

\begin{figure}[h]
	\centering
	\includegraphics[width=\linewidth]{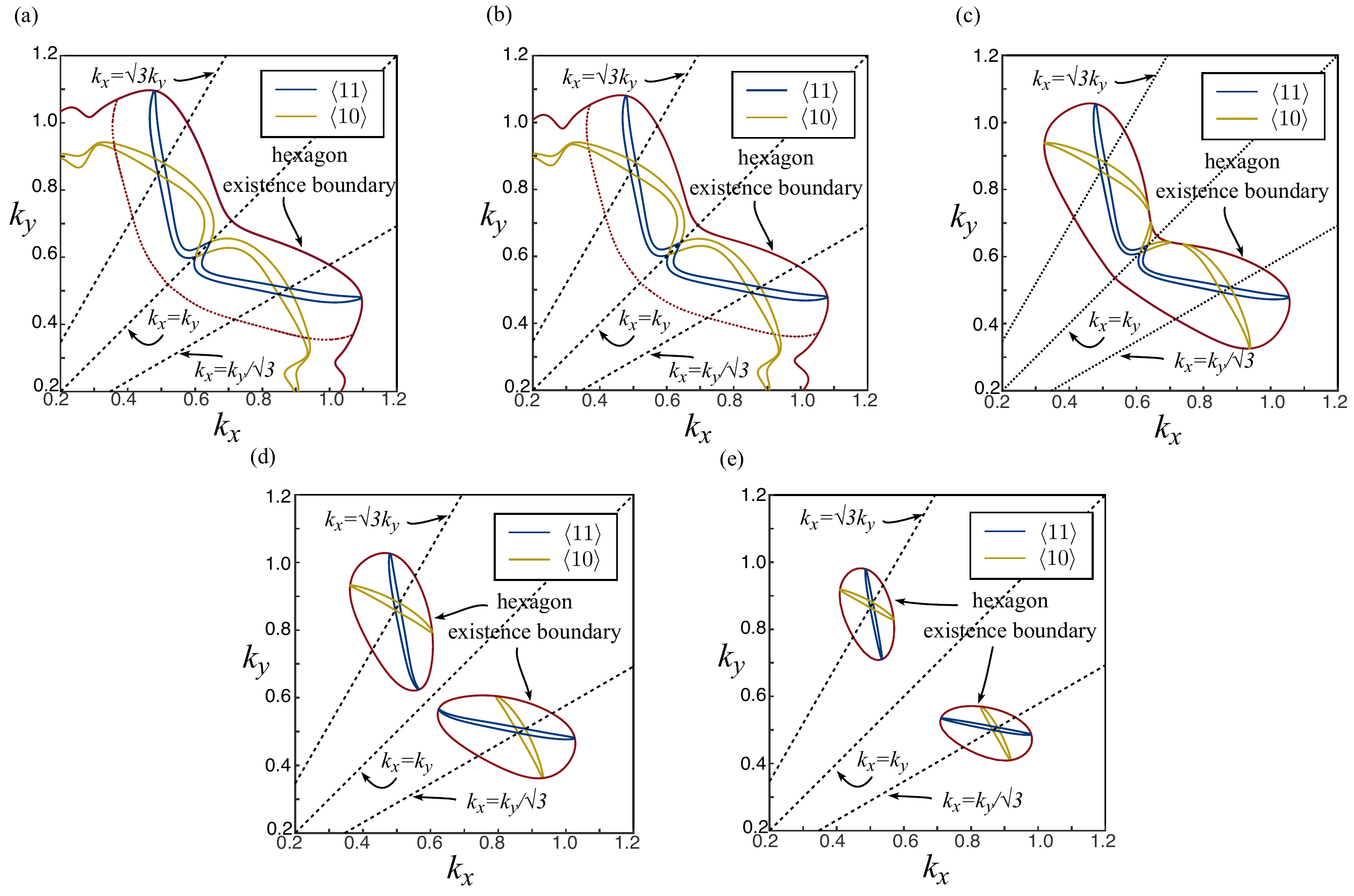}
	\caption{Plots in $(k_x,k_y)$-space of the existence boundary of the distorted hexagons and the Hamiltonian selected distorted $\langle10\rangle$- and $\langle11\rangle$-hexagons for $\nu=1.6$ and (a) $\mu=0.267$, (b) $\mu=0.28$, (c) $\mu=0.3$, (d) $\mu=0.32$, and (e) $\mu=0.345$. We note that the Hamiltonian selected distorted hexagons are always compatible at 4 points; two points along $k_y=\sqrt{3}k_x$ and two points along $k_y = k_x/\sqrt{3}$ lines i.e., perfect hexagons, and in (a)-(c) a further two points along $k_y=k_x$ line i.e., squares.\label{f:stat_compat}}
\end{figure}

In Figure~\ref{f:stat_patch_select}(c) we proposed how a hexagon patch can be viewed as having  $\langle10\rangle$- and $\langle11\rangle$-hexagon fronts perpendicular to each other. From the hexagon front pattern selection criterion~(\ref{e:stat_hex_ham}) discussion above, the $\langle10\rangle$-hexagon front will select a $k_x$ wavenumber for a fixed $k_y$ and similarly the $\langle11\rangle$-hexagon front (now periodic in $x$ and infinite in $y$) will select a $k_y$ wavenumber for a fixed $k_x$. For both these fronts to be compatible in a patch, we then require both the respective selected wavenumbers to intersect and lead to the same selected far-field hexagon cell.  In Figure~\ref{f:stat_compat} we plot the respective selected wavenumbers for the $\langle10\rangle$- and $\langle11\rangle$-fronts for $\nu=1.6$ and $\mu=0.267,0.28,0.3,0.32,0.345$ all of which are contained in the homoclinic snaking region for the $\langle10\rangle$-hexagon front. In~\cite{lloyd2008} it is found that stable 2D stationary patches exist in the homoclinic snaking region for the $\langle10\rangle$-hexagon front, so we take $\mu$-values in this range. 
As shown in Figure~\ref{f:hex_floquet}(b), the hexagon existence boundaries undergoes a significant transition from two `loops' around the linear critical wavenumber values as shown in Figure~\ref{f:stat_compat}(d) and (e) to a single loop in panel (c) to more complicated regions in panels (a) and (b). We also plot the lines where $k_x=\sqrt{3}k_y,k_y$ and $k_y/\sqrt{3}$, which corresponds to perfect hexagons and perfect squares. We note that the diagram should be symmetric about the $k_y=k_x$ diagonal. In all the plots in Figure~\ref{f:stat_compat}, we see 4 intersections of the selected wave numbers that occur on the perfect hexagon lines (2 of these intersections furthest from the origin correspond to stable hexagon fronts whereas the other two intersections correspond to unstable hexagon fronts). In panels (a)-(c), we also see an additional two intersections occurring on the perfect square lines. We expect the square fronts to be unstable to hexagonal perturbations. This computation provides a heuristic reason for the perfect hexagon selection of a patch observed in Lloyd et al.\ \cite{lloyd2008} and in Figure~\ref{f:stat_patch_select}.


\section{Weakly Nonlinear Analysis of Hexagon Fronts}\label{s:weak}


We review the weakly nonlinear analysis of planar hexagon fronts~\cite{malomed1990,doelman2003,escaff2009a,escaff2009b}. Doelman et al.\ \cite{doelman2003} rigorously derived the amplitude equations for $\langle10\rangle$-hexagon invasion fronts and proved existence for fronts in the parameter space $\mu<0$ when the fronts are propagating into an unstable state. 
On a formal level, 
modulated hexagon fronts with different orientations to the underlying hexagonal lattice were considered by~\cite{malomed1990,escaff2009a,escaff2009b}. The leading order amplitude equations are derived by considering the ansatz 
\begin{equation}\label{e:hex_ansatz}
u(x,y,t)= \epsilon \left(\sum_{j=1}^3A_j(\epsilon x,\epsilon^2 t)e^{ik_j\cdot \mathbf{r}} + c.c.\right) + \mathcal{O}(\epsilon^2),
\end{equation}
where $\mathbf{r} = x\hat x + y \hat y$, $\hat x = (\cos\alpha,-\sin\alpha),\hat y = (-\sin\alpha,\cos\alpha)$, $k_1 = (-1,0),k_2=(\frac12,\frac{\sqrt{3}}{2}),k_3=(\frac12,-\frac{\sqrt{3}}{2})$ and $\alpha$ is the orientation of the interface with respect to the hexagonal pattern i.e., $\alpha = 0$ corresponds to a $\langle 10\rangle$-front and $\alpha=\pi/6$ corresponds to a $\langle 11\rangle$-front. Employing the scalings
\[
\mu = \epsilon^2\tilde\mu,\qquad \nu = \epsilon\tilde \nu
\]
and substituting in the ansatz~(\ref{e:hex_ansatz}) in the quadratic-cubic SH equation (\ref{e:sh}), one finds at $\mathcal{O}(\epsilon^3)$ the amplitude equations
\begin{subequations}\label{e:GL_hex}
\begin{align}
(A_1)_T =& M_1\partial_X^2{A}_1 - \left[\tilde\mu + 3\left(|A_1|^2 + 2(|A_2|^2+|A_3|^2) \right) \right]A_1 + 2\tilde\nu \overline{A_2A_3},\\
(A_2)_T =& M_2\partial_X^2{A}_2 - \left[\tilde\mu + 3\left(|A_2|^2 + 2(|A_1|^2+|A_3|^2) \right) \right]A_2 + 2\tilde\nu \overline{A_1A_3},\\
(A_3)_T =& M_3\partial_X^2{A}_3 - \left[\tilde\mu + 3\left(|A_3|^2 + 2(|A_2|^2+|A_1|^2) \right) \right]A_3 + 2\tilde\nu \overline{A_2A_1},
\end{align}
\end{subequations}
where $M_j = 4(k_j\cdot \hat x)^2$ and $(X,T)=(\epsilon x,\epsilon^2 T)$. We look for traveling wave solutions of the form $A(X-\tilde c T)=A(Z)$, where $c=\epsilon\tilde c$.

\begin{figure}[h]
	\centering
	\includegraphics[width=\linewidth]{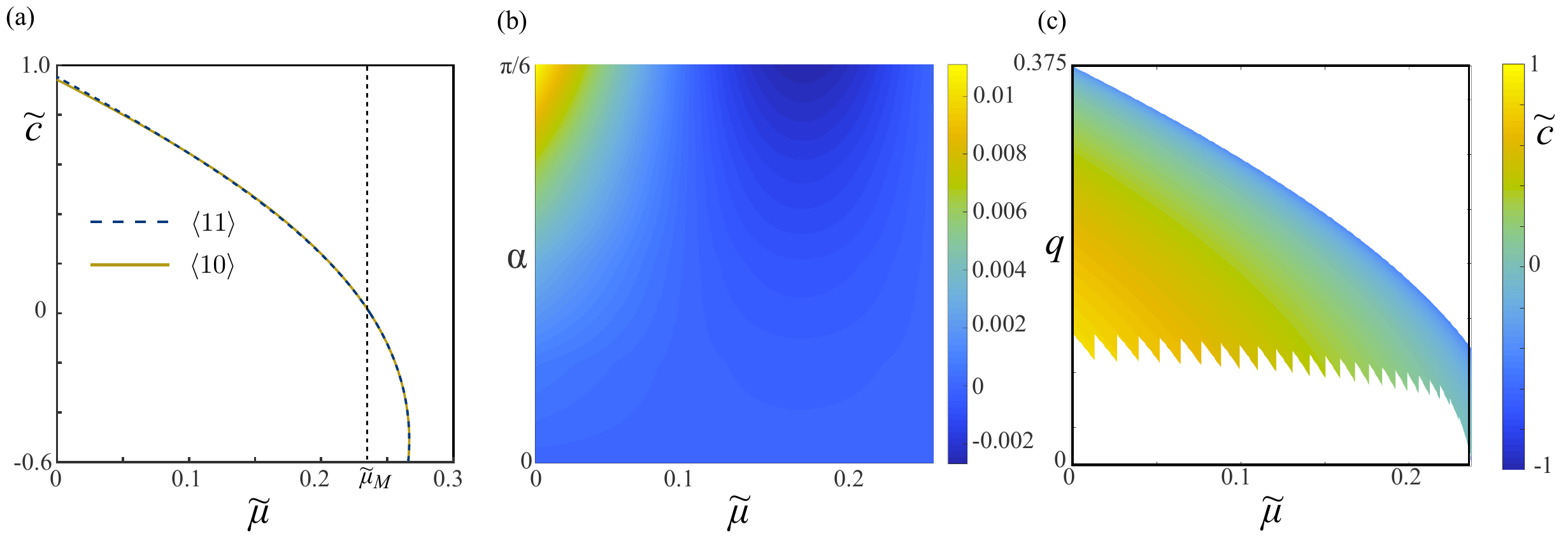}
	\caption{(a) Wave speed of the fronts in the hexagon amplitude equations for $\tilde\nu=2$. (b) the difference in wavespeed of the fronts from the $\langle10\rangle$-hexagon front $\tilde\nu=2$ as the orientation angle $\alpha$ is varied. The Maxwell point i.e., where the fronts become stationary occurs at $\tilde\mu=\tilde\mu_M=0.237$. (c) Existence region of the hexagon fronts connecting to distorted hexagons in the amplitude equations (\ref{e:GL_hex}) with $(A_1,A_2,A_3)=\tilde A(e^{iqX},e^{-iqX/2},e^{-iqX/2}),\tilde A\in\mathbb{R}$ for $\alpha=0$.\label{f:hex_amp}}
\end{figure}

Existence of fronts in~(\ref{e:GL_hex}) is not known for general values of $\tilde c$. For large values of $\tilde c$, singular limit analysis (Doelman et al.\ \cite{doelman2003}) can be used to prove the existence of invading hexagon fronts into the trivial state as well as a multitude of other traveling wave states that we do not investigate here. 
For moderate values of $\tilde c$, we use numerical continuation to map out the selected front speed in $(\tilde\mu,\tilde\nu)$-space; see Figure~\ref{f:hex_amp}. In Figure~\ref{f:hex_amp}(a), we plot the selected wave speed for the $\langle10\rangle$- and $\langle11\rangle$-hexagon fronts for $\tilde\nu=2$. The point where the fronts become stationary (known as the Maxwell point) occurs for $\tilde\nu=2$ at $\tilde\mu=\tilde\mu_M=8\tilde\nu^2/135=0.237$. For $\tilde\mu>\tilde\mu_M$, the fronts have a negative propagation speed and correspond to retreating hexagon fronts. We find that the wave speeds for both the $\langle10\rangle$- and $\langle11\rangle$-hexagon fronts are very similar. In order to observe the difference in the front wave speeds, we plot in Figure~\ref{f:hex_amp}(b), the difference of the wave speeds from the $\langle10\rangle$-invasion front for all values of $\alpha$. Here we observe that the fronts close to the $\langle11\rangle$-invasion front can travel faster or slower than the $\langle10\rangle$-front. In particular, the $\langle11\rangle$-front is slower than the $\langle10\rangle$-front around $\tilde\mu\approx0.16$ and then becomes faster around $\tilde\mu=0$. We note that for $\tilde c=0$, the existence of stationary fronts have been proven using rigorous numerics; see~\cite{berg2015}.

The results here concern fronts connecting to perfect hexagons. It is easy to extend the results here to consider distorted hexagons with off-critical wavenumber in the $X$-direction and critical wavenumber in the $y$-direction by considering $(A_1,A_2,A_3)=(\tilde A_1(Z)e^{iqX},\tilde A_2(Z)e^{-iqX/2},\tilde A_3(Z)e^{-iqX/2})$ which recovers equation~(\ref{e:GL_hex}) in the traveling-frame with the amplitudes replaced by $\tilde A_1,\tilde A_2$ and $\tilde A_3$, and $\mu$ replaced by $\mu-M_jq^2$ in each of the respective amplitude equations. Due to the small dependence of the front speed on $\alpha$, we do not trace out the solution space in $\alpha$. However, around each perfect hexagon front branch is an envelope of distorted hexagon fronts for $q\in[0,q_c(\mu)]$ where $q_c(\mu)=0$ for $\mu$ at the fold of the perfect hexagons in the bistable region and $q_c$ grows monotonically as $\mu$ is decreased; see Figure~\ref{f:hex_amp}(c). We note that additional equivariant terms~\cite{hoyle2006} in the hexagon amplitude equations may select the wavenumber similar to that seen in the 1D amplitude equation~\cite{lloyd2019} at this order.

\section{Numerical Method for Invasion Fronts}\label{s:method}
In this section, we describe the numerical method for time simulations, and the numerical boundary value problem for the invasion fronts. 

\subsection{Initial Value Solver}\label{s:ivp_method}
Time simulations of~(\ref{e:sh}) are carried out on a periodic rectangle using a 4th order exponential time-stepper Runge-Kutta scheme in time and fast Fourier method in space~\cite{kassam2003,kassam2005}. The scheme is implemented in \textsc{Matlab2017b} on a dual hexa-core (2.93 GHz) Mac Pro with 24GB RAM. 

\subsection{Hexagon Invasion Fronts}
We describe the far-field core decomposition method for computing hexagon invasion fronts. This method is very similar to that described in Lloyd~\cite{lloyd2019}. 
We first introduce the following change of coordinates $(z,\tilde y,\tilde t) = (x-k_t t/k_x,k_y y,k_t t)$ so that $(z,\tilde y,\tilde t)\in\mathbb{R}\times S^1\times S^1$ where $S^1 = \mathbb{R}/2\pi\mathbb{Z}$. We truncate the domain in $z$ to $[-L_z,L_z]$. Dropping tildes, the SH equation becomes
\begin{equation}\label{e:t2SH}
0=cu_z-k_t u_t  -(1+\partial_z^2 + k_y\partial_y^2)^2u - \mu u + F(u) =: Lu + F(u),
\end{equation}
where 
\[c=\frac{k_t}{k_x}.
\]
 We consider the far-field core decomposition of the fronts given by,
\begin{equation}\label{e:far_core}
u(z,y,t) = u_h(k_xz + t,y;k_x)\chi(z) + w(z,y,t),
\end{equation}
where $u_h(\xi,y)$ is the domain-covering hexagon cell and solves the stationary 2D Swift-Hohenberg equation
\begin{equation}\label{e:hex_SH}
0 = -(1+k_x^2\partial_x^2 + k_y^2\partial_y^2)^2u_h + \mu u_r + F(u_h),
\end{equation}
on the domain $(x,y)\in [0,2\pi)^2$, and we impose Neumann boundary conditions. The function $\chi(z)=1-\frac12[1+\mbox{tanh}(m(z-d))]$ is a smooth function going to unity for $z<d$ and zero when $z>d$. The `core' function $w(z,y,t)$ describes the interface and decays to zero as $|z|\rightarrow\infty$. Substituting~\eqref{e:far_core} into~\eqref{e:t2SH} yields the inhomogeneous PDE for $w$ given by
\begin{equation}\label{e:PDE_w}
L[u_h\chi + w] + F(u_h\chi + w) - \chi\left( Lu_h + F(u_h)\right) = 0,
\end{equation}
where the far-field hexagon equation $\chi\left( Lu_h + F(u_h)\right)=0$ has been subtracted. We add two additional phase conditions:
\begin{subequations}\label{e:phase_cond}
\begin{align}
\int_0^{2\pi}\int_0^{2\pi}\int_{-L_z}^{L_z}([c\partial_z - k_t\partial_t ]w^{\mbox{old}}_z)(w-w^{\mbox{old}})dzdydt =& 0,\\ \int_0^{2\pi}\int_0^{2\pi}\int_{-L_z}^{-L_z+2\pi/k_x}(\partial_xu_h)wdzdydt =& 0,
\end{align}
\end{subequations}
where the template function $w^{\mbox{old}}$ is a previously computed solution. The first phase condition is the standard condition used to select the wavespeed of fronts $c$~\cite{krauskopf2007}, while the second phase condition is the standard condition to select the far-field spatial wavenumber $k_x$~\cite{lloyd2017}. We note that no additional phase condition is required for translations in $y$ as the interpolated far-field hexagons break this symmetry in~(\ref{e:PDE_w}). 

The PDE~(\ref{e:PDE_w}) is discretized with fourth-order finite differences in $z$, and pseudo-spectral Fourier method in $y$ and $t$; see~\cite{lloyd2019}. We use the trapezoidal rule to approximate the phase conditions. The hexagon cell equation~(\ref{e:hex_SH}) is discretized using a pseudo-spectral Fourier method in both $x$ and $y$ and is interpolated using a band limited interpolant~\cite{trefethen2000}. The discretized form of the system~(\ref{e:PDE_w})-(\ref{e:phase_cond}) form a large algebraic nonlinear system that is solved using Newton's method. A single Newton step requires solving a large bordered system and to solve it we use a version of the bordering method by Phipps and Salinger~\cite{phipps2006} that is explained in the Appendix. 
We modify \textsc{Matlab}'s \verb1fsolve1 routine\footnote{The GitHub code repository linked at the end of this paper combines the bordering method explained in the appendix with a Newton Amijo method~\cite{kelley2003}} to use the bordering algorithm. This algorithm employs a trust-region based method that helps convergence to a solution despite the initial guess for the Newton method having a large residual. The boundary value problem is then embedded into a numerical continuation routine~\cite{avitabile2020} to path follow the solutions as parameters are varied. 

\begin{figure}[p]
	\centering
	\includegraphics[width=\linewidth]{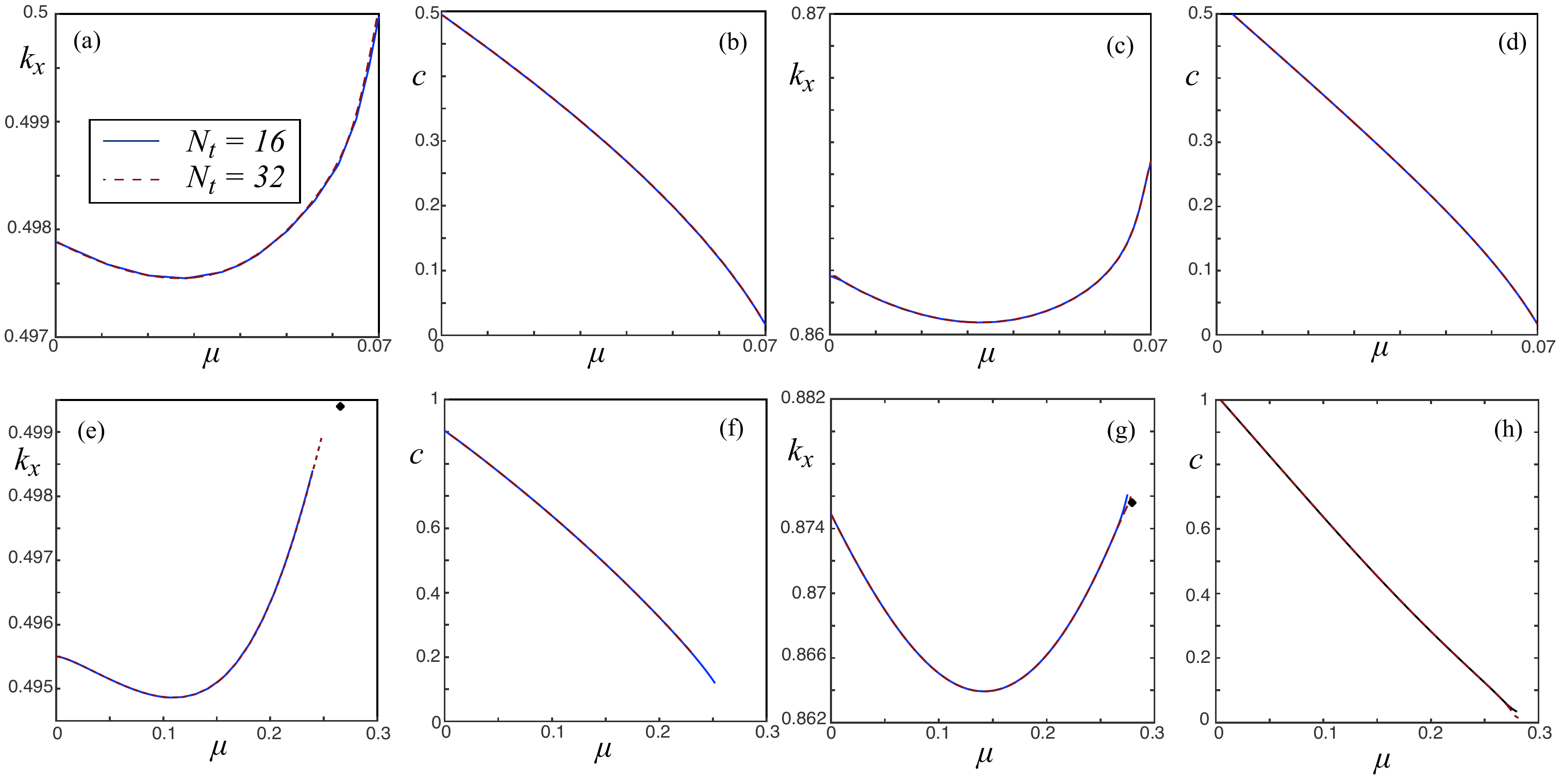}
	\caption{Plots of the selected $(k_x,c)$ for the hexagon invasion fronts with numerical discretizations $N_z=400,N_y=16,m=1,d=40,z\in[-50\pi,10\pi]$ and $N_t=16,32$. Panels (a)-(b) show the selected $(k_x,c)$ for the $\langle10\rangle$-front $\nu=0.9,k_y=0.86$, (c)-(d) panels show the selected $(k_x,c)$ for the $\langle11\rangle$-front $\nu=0.9,k_y=0.5$, panels (e)-(f) show the selected $(k_x,c)$ for the $\langle10\rangle$-front $\nu=1.6,k_y=0.8$, and panels (g)-(h) show show the selected $(k_x,c)$ for the $\langle11\rangle$-front $\nu=1.6,k_y=0.45$. The black dots in (e) and (g) correspond to the Hamiltonian selected wavenumber for the hexagons on the edge of the snaking region.\label{f:hex_converge}}
\end{figure}

We use the following initial condition
\[
w(z,y,t) = (H(z)-H(z-d))u_h(k_xz+t,y;k_x),
\]
where $H$ is the Heaviside function with $(k_x,k_y)=(0.5,0.86)$ for the $\langle10\rangle$-hexagon fronts and $(k_x,k_y)=(0.86,0.5)$ for the $\langle11\rangle$-hexagon fronts. Sometimes convergence from these initial conditions is poor and quicker convergence is found if we let $k_y$ also be solved for with the additional phase condition
\[
\int_0^{2\pi}\int_{L_x}^{L_x}([k_y\partial_y - k_t\partial_t ]w^{\mbox{old}}_z)(w-w^{\mbox{old}})dzdt = 0.
\]
Once a solution is converged, the continuation is then done with $k_y$ fixed and this phase condition turned off. 

\begin{figure}[p]
	\centering
	\includegraphics[width=\linewidth]{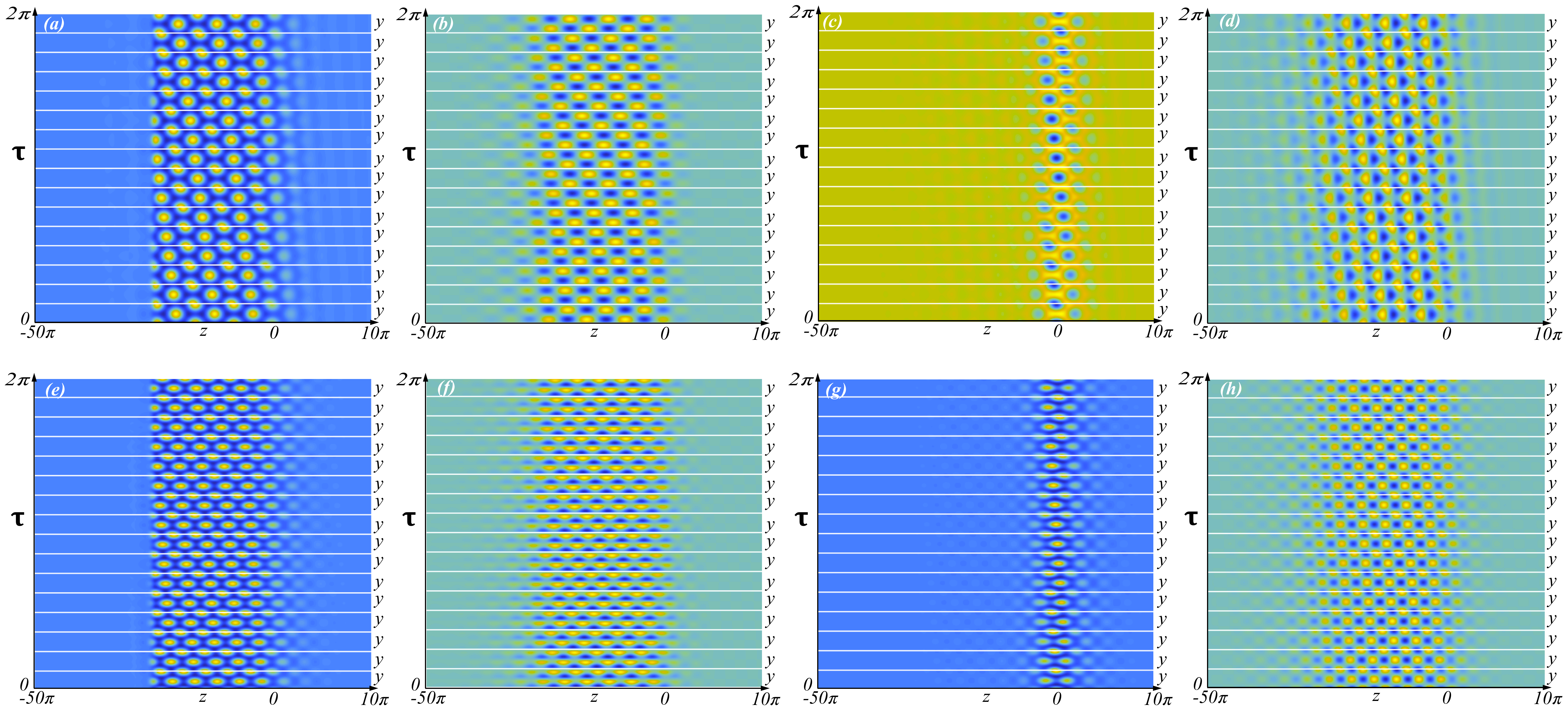}
	\caption{(a) profile of the core function $w$ for the $\langle10\rangle$-front with $(\mu,\nu,k_y)=(0.04,0.9,0.86)$ and corresponding approximate neutral modes (b) $w_y$, (c) $[c\partial_z - k_t\partial_t ]w$, (d) $w_z$. (e) profile of the core function $w$ for the $\langle11\rangle$-front with $(\mu,\nu,k_y)=(0.04,0.9,0.5)$ and corresponding approximate neutral modes (f) $w_y$ (g) $[c\partial_z - k_t\partial_t ]w$ (h) $w_z$. The numerical discretization is $N_z=400,N_y=16,N_t=16$ for all plots. Both the fronts are found to be stable with respect to co-periodic perturbations.\label{f:zeroeigs}}
\end{figure}

We note that stationary hexagon fronts can be computed in exactly the same way with $c=0$ and only the phase condition (\ref{e:phase_cond}b). This is particularly useful if there is no spatial Hamiltonian to help compute the selected far field wavenumbers in general reaction-diffusion systems.

Typical discretizations used for the invasion fronts are $N_z=400,N_y=16,N_t=16$ with $m=1,d=40,z\in[-50\pi,10\pi]$. It is anticipated that as one approaches the homoclinic snaking region and the front speed tends to zero, that the convergence in $t$ will become worse (as expected for any numerical scheme). In Figure~\ref{f:hex_converge}, we plot the selected $(k_x,c)$ for the hexagon invasion fronts with $N_t=16$ and $32$. We find for both $\nu=0.9$ and $\nu=1.6$ that there is good convergence of the hexagon fronts for $N_t=16$ for the entire range of $\mu$ values up to the homoclinic snaking region. In particular, the selected spatial wavenumber of the invasion fronts converges to that of the corresponding snaking front at the edge of the snaking region. 
We have tested the algorithm with different discretizations for the other computational parameters and the results do not significantly change; see also~\cite{lloyd2019}. 

Finally, we plot in Figure~\ref{f:zeroeigs} typical profiles of the core function for the $\langle10\rangle$- and $\langle11\rangle$-fronts in $(z,y,t)$-space, and corresponding neutral modes computed using \textsc{Matlab}'s \verb1eigs1 function for the discretized linearization of (\ref{e:PDE_w}). It is easy to show that the linearization of the fronts should have the neutral modes, $w_y, [c\partial_z - k_t\partial_t ]w$, and $w_z$; we find precisely these modes with corresponding eigenvalues $\mathcal{O}(10^{-4})$. Both the fronts in Figure~\ref{f:zeroeigs} are found to be stable with respect to co-periodic perturbations.


\section{Results: Hexagon Invasion Fronts}\label{s:numerics}

We present our main results for hexagon invasion fronts. Lloyd et al.\ \cite{lloyd2008} looked at stationary hexagon fronts in the two cases of $\nu=0.9$ and $1.6$ corresponding to small and moderate hysteresis in the SH equation. To be consistent, we chose the same $\nu$ values. We then look at hexagon invasion fronts in a non-variational version of the Swift-Hohenberg equation. 

\subsection{Small Hysteresis Case: $\nu=0.9$}

\begin{figure}[h]
	\centering
	\includegraphics[width=0.8\linewidth]{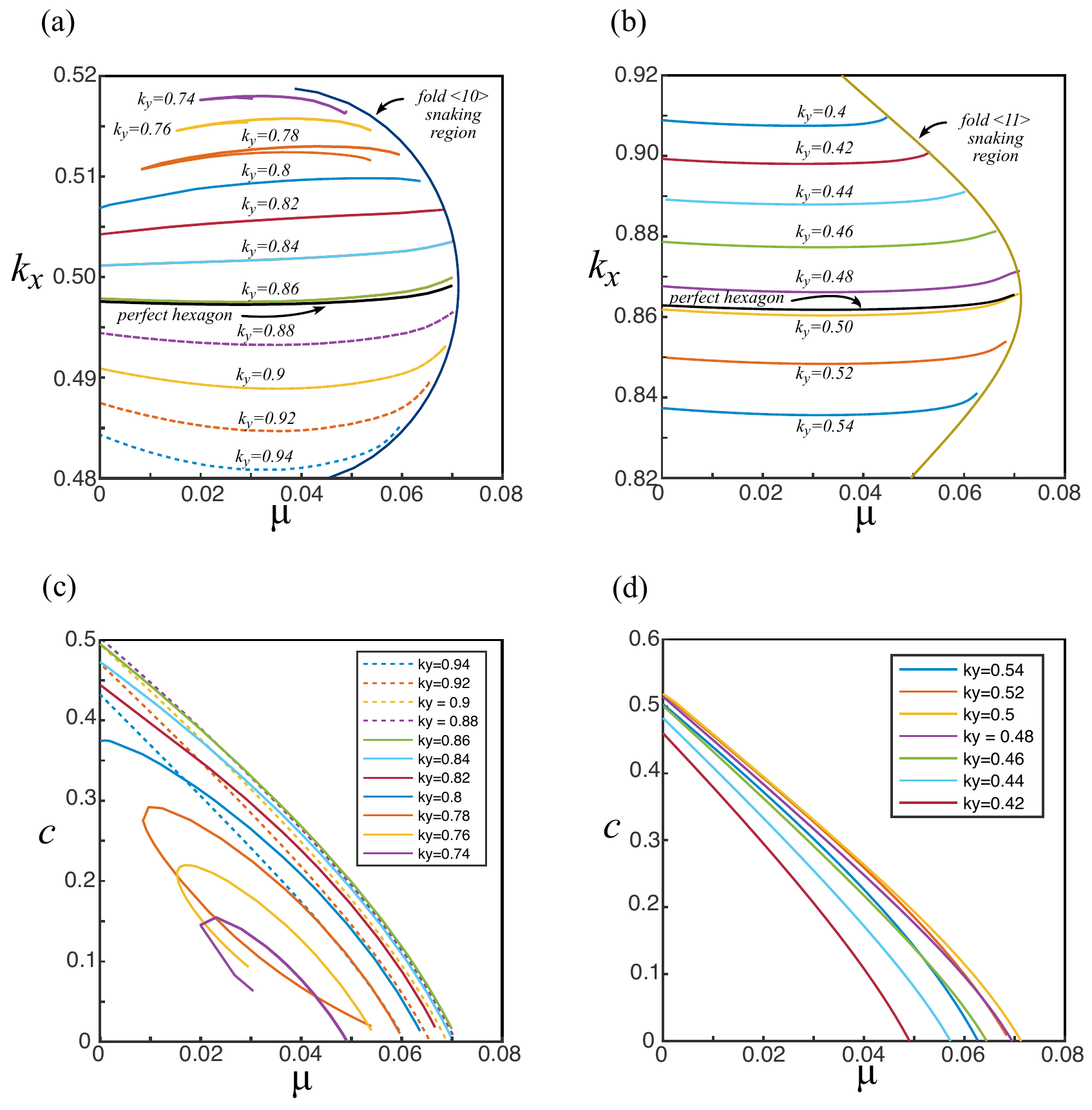}
	\caption{The selected wavenumbers for (a) $\langle10\rangle$-front (b) $\langle11\rangle$-front and corresponding invasion speeds in (c) and (d) in the SH equation with $\nu=0.9$.\label{f:hex_invade_nu_0_9}}
\end{figure}

For $0<\nu<1.049$, the snaking regions of the hexagon fronts become very narrow, almost overlapping, and are difficult to continue. However, the invasion fronts exist in a larger region in parameter space making them easy to continue. In Figure \ref{f:hex_invade_nu_0_9}, we show the selected far-field wavenumber and front speed for both the $\langle10\rangle$- and $\langle11\rangle$-fronts. For the $\langle10\rangle$-front, we see that fronts with $k_y$ close to $\sqrt{3}/2$ (including the $\langle10\rangle$-front with perfect hexagons) starts at the edge of the homoclinic snaking region with the Hamiltonian selected wavenumber and ``dips'' down (leading to stretched hexagons in $x$) before rising again as $\mu$ is decreased leading to compressed hexagons in $y$. However, for smaller $k_y$'s less than $0.8$, the $\langle10\rangle$-front selected $k_x$ instead has a ``hump'' as $\mu$ decreases where the hexagon is initially compressed then stretched. We also plot for clarity the wavenumber selection for fronts that select the perfect hexagon and see that this tracks very closely the $k_y=0.86$ front for the $\langle10\rangle$-front and $k_y=0.5$ front for the $\langle11\rangle$-front i.e., close to the linearly critical wavenumbers.

\begin{figure}[h]
	\centering
	\includegraphics[width=\linewidth]{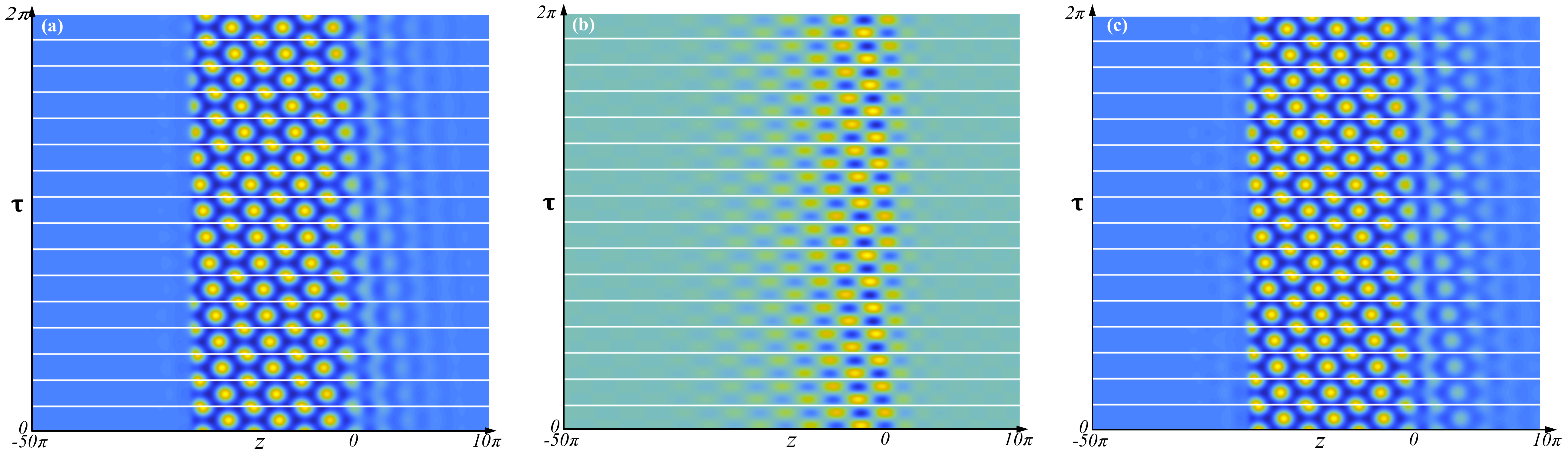}
	\caption{Plots of the $\langle10\rangle$-front and localized neutral mode at the fold ($(\nu,k_y)=(0.9,0.76)$) with panel (a) showing the core profile $w$ at $\mu=0.015$, and panel (b) the corresponding additional localized neutral mode. Panel (c) shows a plot of the $\langle10\rangle$-front core profile, $w$, past the fold at $(\mu,\nu,k_y)=(0.01,0.9,0.78)$.\label{f:fold_eig}}
\end{figure}

\begin{figure}[h]
	\centering
	\includegraphics[width=\linewidth]{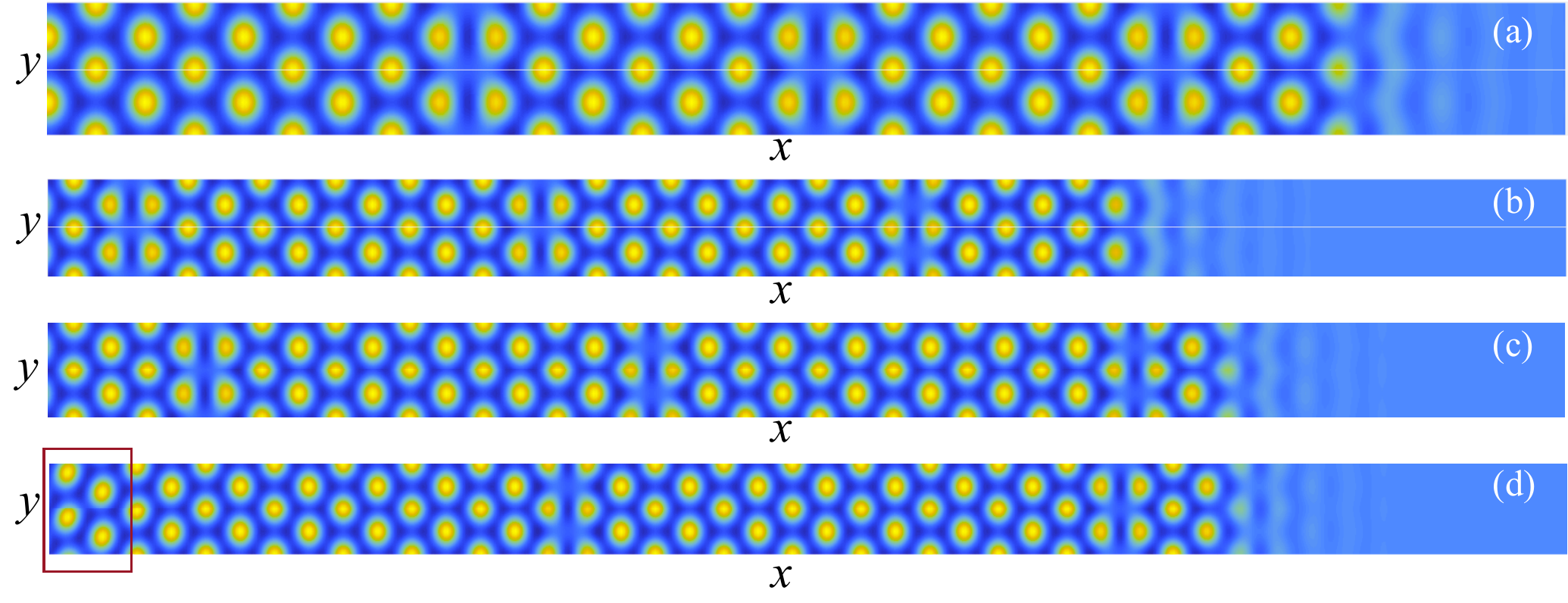}
	\caption{Plots of $\langle10\rangle$-fronts with defects from time simulations of (\ref{e:sh}) with $(\nu,k_y)=(0.9,0.76)$ and (a) $\mu=0.007$ (b) $\mu=0.012$ (c) $\mu=0.013$ (d) $\mu=0.014$. The profiles have been periodically extended once in $y$. In the red box in panel (d), we see how the defects slowly shear back to becoming a regular cellular hexagon lattice.\label{f:front_defect_ivp}}
\end{figure}

For $k_y$ less than $0.8$, we find that the $\langle10\rangle$-front bifurcation branch undergoes a fold for small $\mu$. In Figure~\ref{f:fold_eig}, the $\langle10\rangle$-front and corresponding additional localized neutral mode at the fold for $(\mu,\nu,k_y)=(0.015,0.9,0.76)$. This localized neutral mode suggests that periodic ``square'' defects should occur past the fold. We also plot the core function after we have passed around the fold showing what looks like a ``defect'' in the growth of the front and the front speed tends to zero. In Figure~\ref{f:front_defect_ivp}, we plot a few time simulations of the $\langle10\rangle$-fronts with defects for $(\nu,k_y)=(0.9,0.76)$ as we approach the fold from below at $\mu\approx0.015$. As expected, we observe that the time gaps between defects to increase as we approach the fold. It appears the defects are not perfectly periodic in $x$ and occasionally occur one column of cells earlier than expected suggesting a spatial quasi-periodic structure. We note that the defects are unstable, and they deform back to distorted hexagons (via shearing) as shown in Figure~\ref{f:front_defect_ivp}(d) in the red box. In principle the numerical continuation method for hexagon invasion fronts should be able to compute the $\langle10\rangle$-fronts with periodic defects but due to the size of the computation requires we have been unable to do this. 

\begin{figure}[h]
	\centering
	\includegraphics[width=\linewidth]{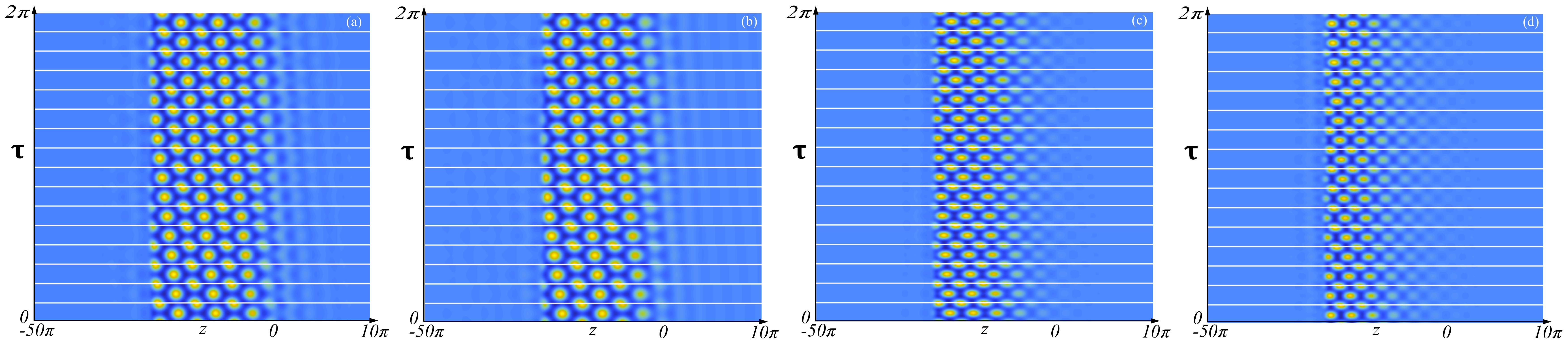}
	\caption{Plots of the core function, $w$, of the $\langle10\rangle$-fronts with $(\mu,\nu)=(0.04,0.9)$ and (a) $(k_y,k_x,k_t) = (0.7389, 0.5178, 0.0385)$ (b) $(\mu,k_y,k_x,k_t) = (0.9766, 0.4768, 0.0302)$ and plots of the core function of the $\langle11\rangle$-fronts with $(\mu,\nu)=(0.04,0.9)$ (c) $(\mu,k_y,k_x,k_t) = (0.5624, 0.8205, 0.1190)$ and (d) $(\mu,k_y,k_x,k_t) = (0.4022, 0.9122, 0.0257)$.\label{f:profiles10nu09}}
\end{figure}

For $k_y<0.74$ and $k_y>0.94$, we observe detachment of the $\langle10\rangle$-front from the hexagons where the front propagates too fast; see Figure~\ref{f:profiles10nu09}(a) and (b). This is likely to be occurring due to the close presence of the existence boundaries for the hexagons. 

The wavenumber selection for the $\langle11\rangle$-fronts is shown in Figure \ref{f:hex_invade_nu_0_9}(b) and (d). Here we observe that we can continue all branches up to the edge of the homoclinic snaking region and that they observe the characteristic ``dip'' in their selected wavenumber as seen in the 1D case~\cite{lloyd2019}. For $k_y<0.4$ and $k_y>0.54$, we find that detachment of the front from the hexagons where the front propagates too fast as we get close to the existence boundaries of the hexagon cells; see Figure~\ref{f:profiles10nu09}(c) and (d). Again, we also see that the wavenumber selection of the fronts that select the perfect hexagon tracks very closely the linearly critical wavenumber i.e., $k_y=0.5$. 

We show front speeds for both the  $\langle10\rangle$- and $\langle11\rangle$-fronts in Figure \ref{f:hex_invade_nu_0_9}(c) and (d). It is found that the fastest fronts are those with $k_y$ chosen to be around the linear critical hexagon wavenumber. All the front speeds tend to zero as one approaches the edge of the snaking region. In Figures~\ref{f:hex_invade_nu_0_9}(c) and (d), we observe that the $\langle11\rangle$-front is slightly quicker than the $\langle10\rangle$-front for $\mu$ near zero and the $\langle10\rangle$-front is slightly quicker than the $\langle11\rangle$-front as $\mu$ gets closer to the snaking region. This matches with the weakly nonlinear asymptotics in \ref{f:hex_amp}(b) with $\alpha = \pi/6$ corresponding to the $\langle11\rangle$-front.

\begin{figure}[h]
	\centering
	\includegraphics[width=0.8\linewidth]{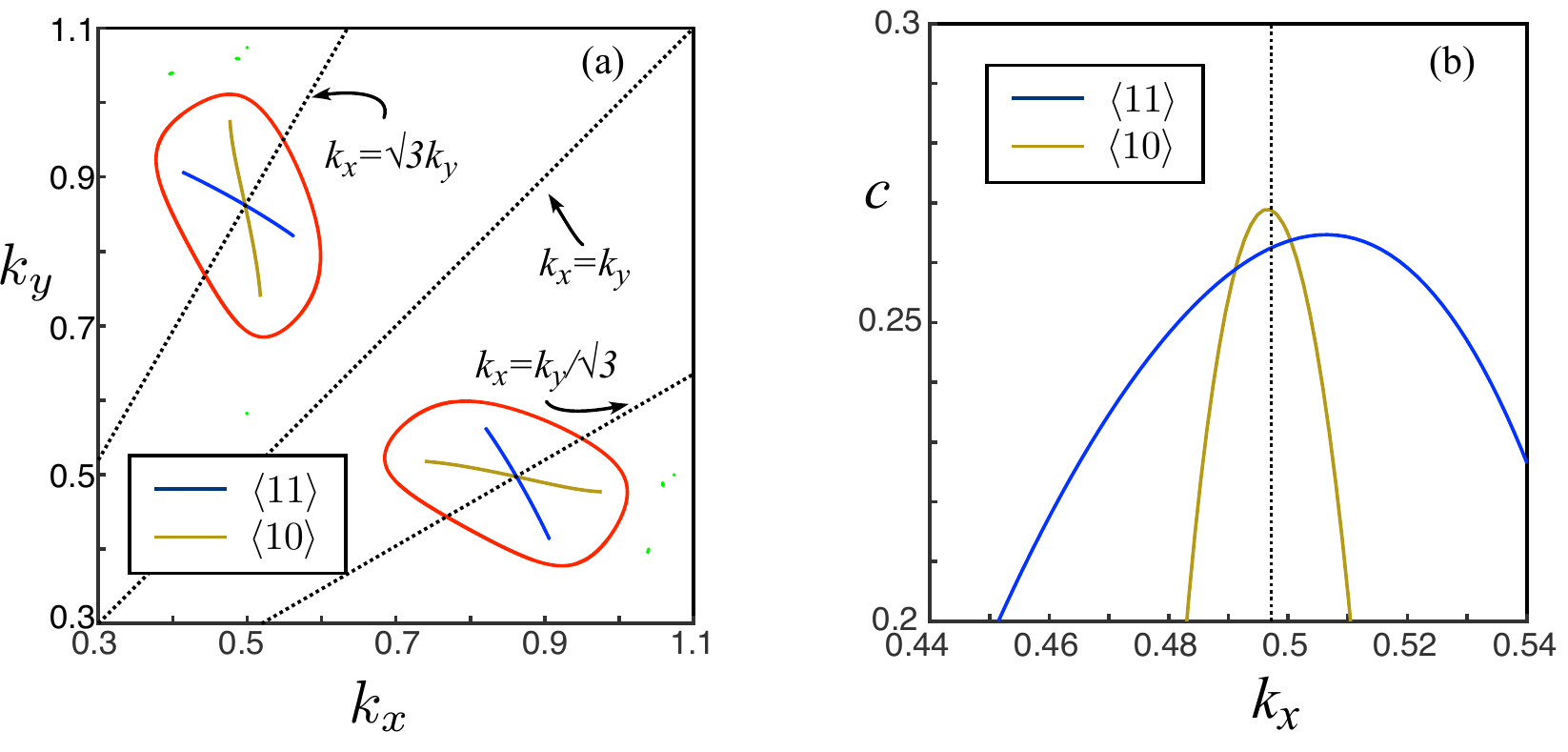}
	\caption{The compatibility diagram of the $\langle10\rangle$- and $\langle11\rangle$-fronts with $(\mu,\nu)=(0.04,0.9)$ and panel (a) the selected far-field wavenumbers in $(k_x,k_y)$-space, and (b) the selected front speeds. The vertical dashed line in (b) denotes the hexagon compatible wavenumber.\label{f:comp_hex_nu_0_9}}
\end{figure}

We next plot the compatibility diagrams for the $\langle10\rangle$- and $\langle11\rangle$-fronts and corresponding front speeds for $(\mu,\nu)=(0.04,0.9)$ in Figure~\ref{f:comp_hex_nu_0_9}. For these parameter values, the existence regions of the hexagons form closed regions in $(k_x,k_y)$-space. At the ends of the select far-field wavenumbers for the $\langle10\rangle$- and $\langle11\rangle$-fronts (shown in Figure~\ref{f:comp_hex_nu_0_9}(a)) we observe detachment of the front before reaching the existence boundaries of the cellular distorted hexagon pattern. However, we believe these curves would terminate at the existence boundaries if we computed on larger $z$ domains. We observe an intersection of the far-field wavenumbers almost on the perfect hexagon line ($\sim\mathcal{O}(10^{-4})$ off) but this is at the numerical tolerance, and we believe the intersection should occur on the perfect hexagon line. The selected front speed at the intersection of the far-field wavenumbers yields that the $\langle10\rangle$-front should propagate quicker that the $\langle11\rangle$-front. Interestingly, the $\langle11\rangle$-front is predicted to propagate quicker than the $\langle10\rangle$-front except in a small region around $k_x=0.5$.

\subsection{Moderate Hysteresis Case: $\nu=1.6$}

\begin{figure}[htb]
	\centering
	\includegraphics[width=0.96\linewidth]{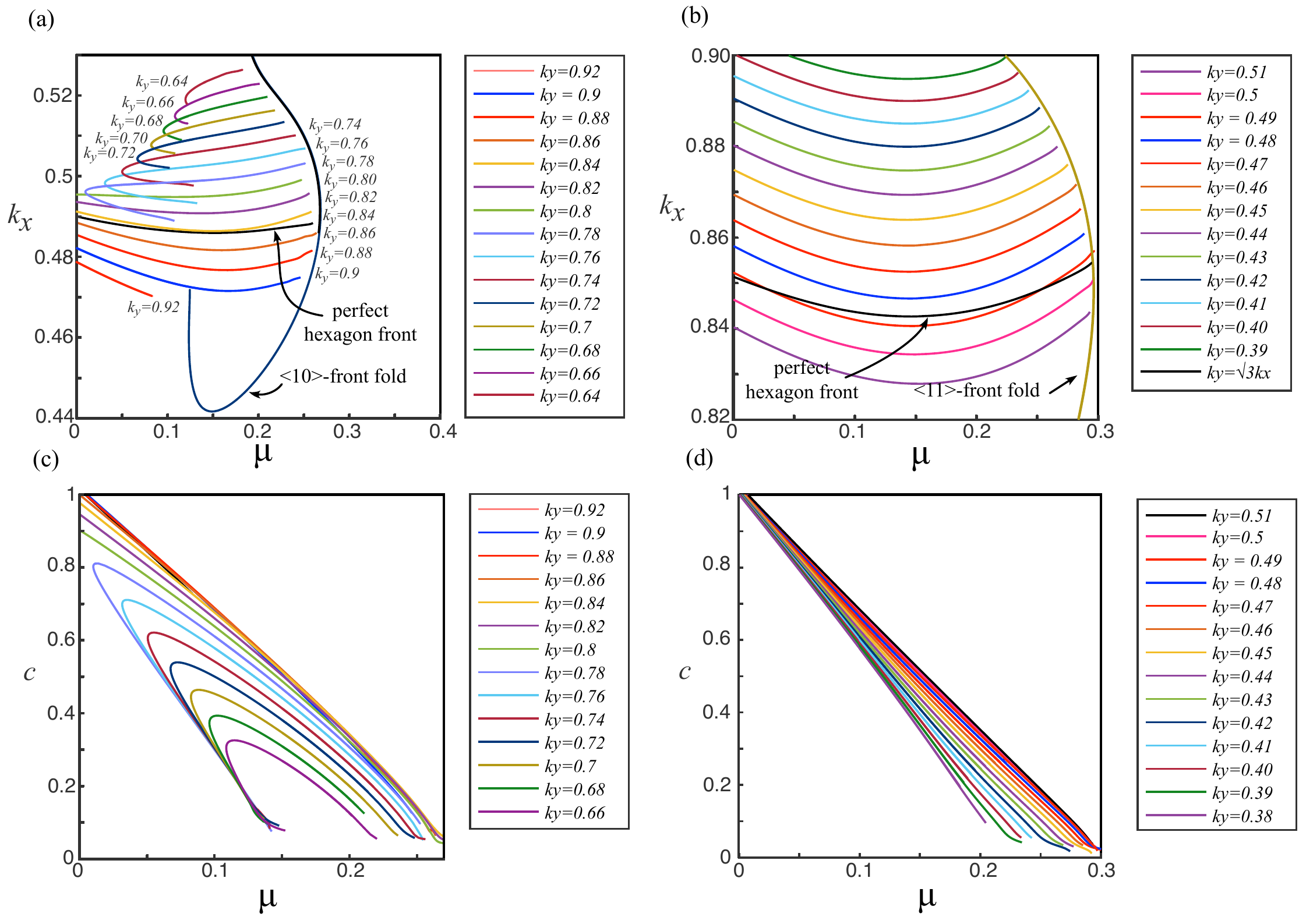}
	\caption{The selected wavenumbers for (a) $\langle10\rangle$-front (b) $\langle11\rangle$-front and corresponding invasion speeds in (c) and (d) of the SH equation with $\nu=1.6$. 
	\label{f:hex_invade}}
\end{figure}

We now investigate hexagon invasion fronts for moderate hysteresis with $\nu=1.6$. In Figure~\ref{f:hex_invade}, we plot the selected wavenumbers and front speeds for both the $\langle10\rangle$- and $\langle11\rangle$-hexagon fronts.

In Figure~\ref{f:hex_invade}(a), we see that the $\langle10\rangle$-fronts for $0.8<k_y<0.9$ start at the edge of the snaking region and display a dip in the selected far-field wavenumber $k_x$ as $\mu$ is decreased. The perfect hexagon front follows very closely the $k_y=0.84$ front. For $0.64<k_y<0.8$, we again see the $\langle10\rangle$-fronts develop a fold for small $\mu$. This fold marks the point where defect forming invasion fronts occur as seen in Figure~\ref{f:stat_hex_snake}(d). 
For $k_y>0.9$, we observe that the $\langle10\rangle$-front detaches from the hexagons due to the front propagating too fast similar to what we observe in the small hysteresis case; see Figure~\ref{f:profiles10nu09}.

 The selected wavenumber of the $\langle11\rangle$-fronts is shown in Figure~\ref{f:hex_invade}(b). Here we observe that the fronts do not have any folds in $\mu$ unlike the $\langle10\rangle$-fronts. A key difference from the small hysteresis case is that the $\langle11\rangle$-invasion fronts exist, as expected, significantly beyond edge of the $\langle10\rangle$-snaking fronts up to the edge of the $\langle11\rangle$-snaking fronts. 
 
\begin{figure}[htb]
	\centering
	\includegraphics[width=0.8\linewidth]{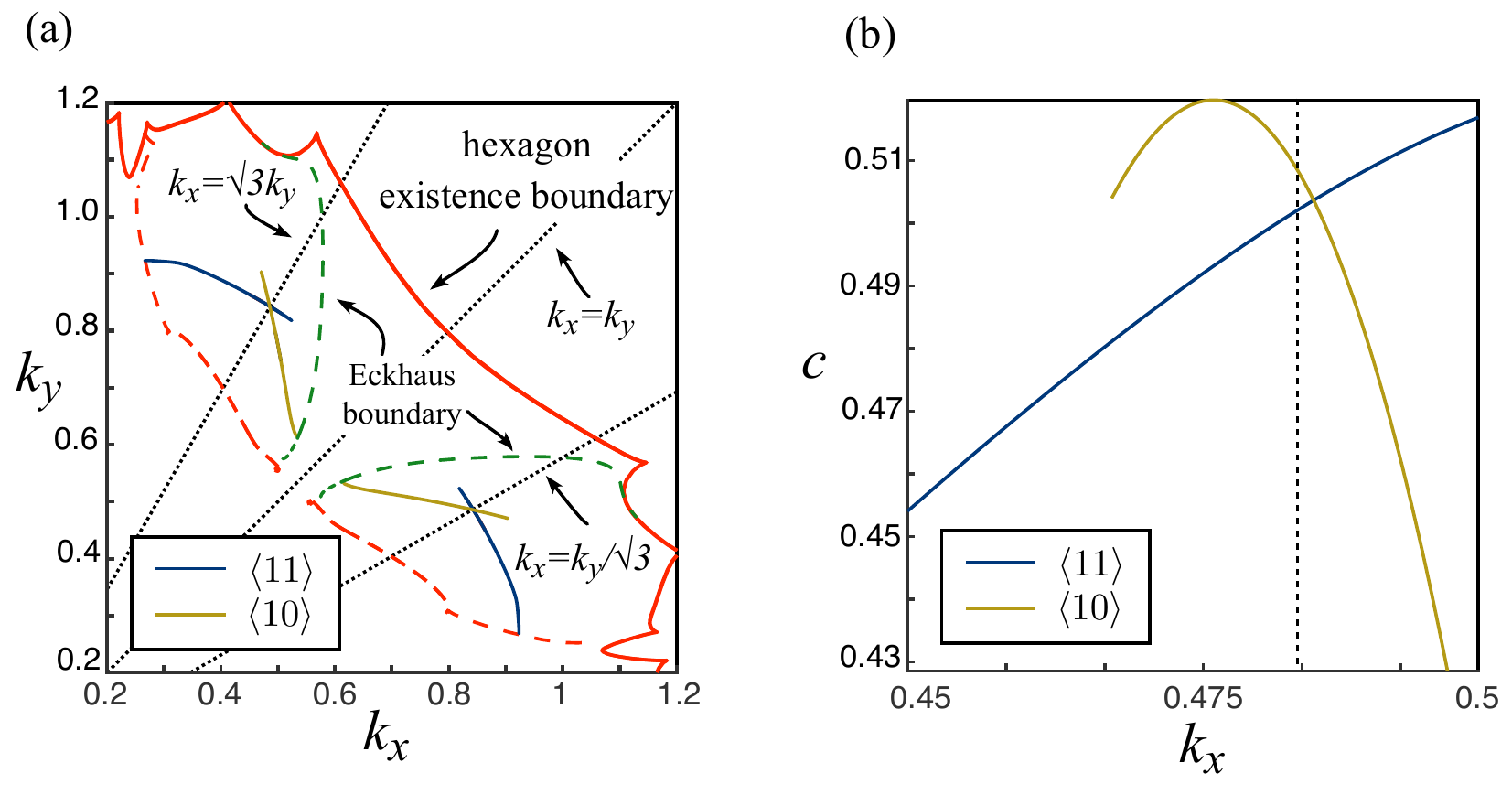}
	\caption{The compatibility diagram of the $\langle10\rangle$- and $\langle11\rangle$-fronts with $(\mu,\nu)=(0.15,1.6)$ and panel (a) the selected far-field wavenumbers in $(k_x,k_y)$-space, and (b) the selected front speeds. The vertical dashed line in (b) denotes the hexagon compatible wavenumber that occurs at $(k_x,k_y)=(0.4857,0.8433)$.\label{f:hex_compat_nu_1_6}}
\end{figure}

In Figure~\ref{f:hex_compat_nu_1_6}, we plot the compatibility diagram for the $\langle10\rangle$- and $\langle11\rangle$-fronts with $(\mu,\nu)=(0.15,1.6)$. We again find that the only compatible hexagon is the perfect one. For the $\langle10\rangle$-front, we see one end of the branch terminates on the Eckhaus stability boundary whereas the other end of the $\langle11\rangle$-front branch terminates on the hexagon co-periodic stability boundary. The other ends of the front branches are terminated due to detachment as discussed above. The selected front speeds for the compatible hexagon fronts again show that the $\langle10\rangle$-front should propagate slightly faster than the $\langle11\rangle$-front. 

\subsection{Non-Variational Swift-Hohenberg Equation}

\begin{figure}[htb]
	\centering
	\includegraphics[width=0.8\linewidth]{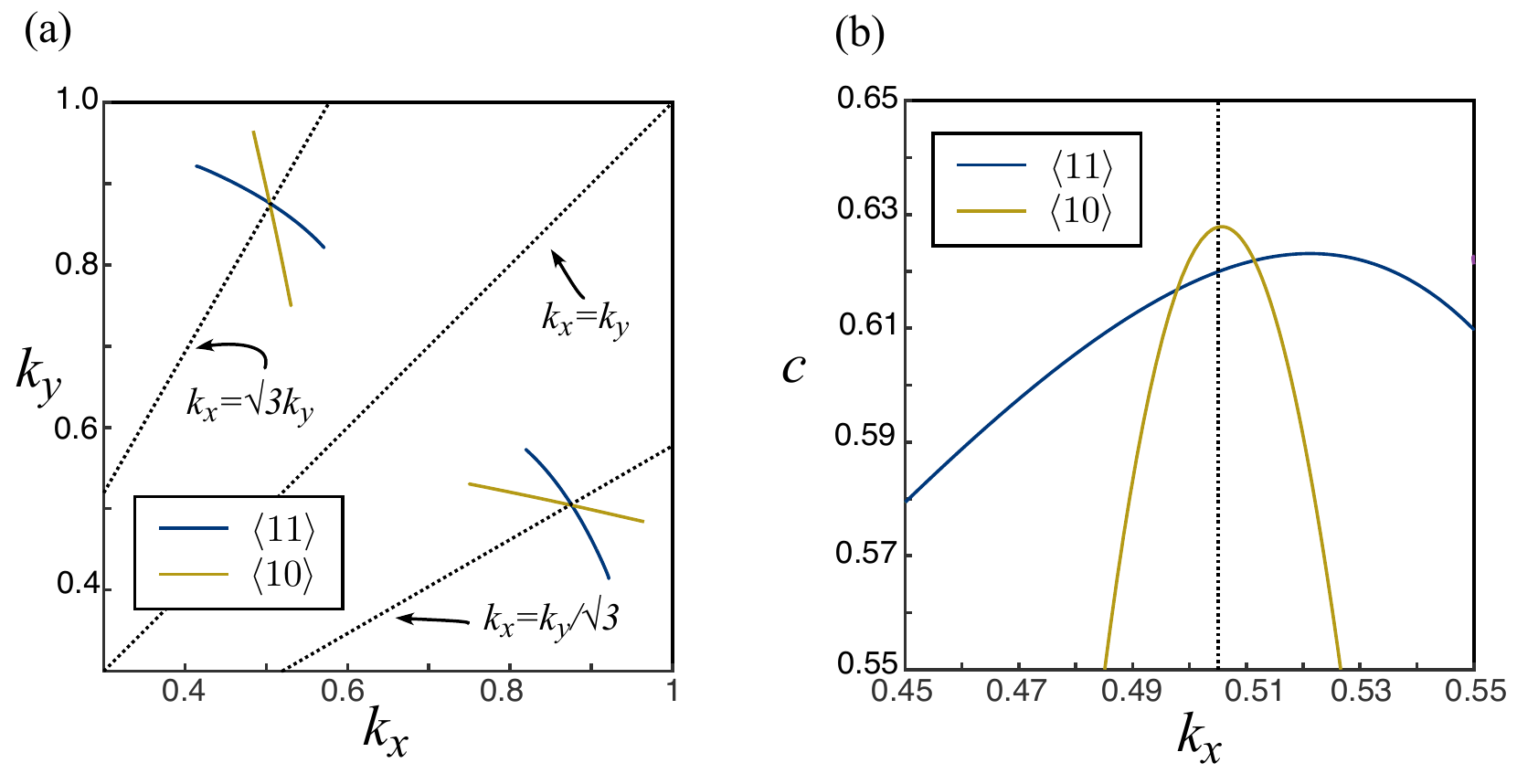}
	\caption{Compatibility diagram of the $\langle10\rangle$- and $\langle11\rangle$-fronts with $(\mu,\nu,\beta)=(0.15,2,-0.5)$ in the non-variational SH equation~(\ref{e:non_SH}) (a) the selected far-field wavenumbers in $(k_x,k_y)$-space and (b) the selected front speeds. The vertical dashed line in (b) denotes the hexagon compatible wavenumber that occurs at $(k_x,k_y) =(0.5041,0.8760)$.
\label{f:non_var_comp}}
\end{figure}

We next demonstrate the compatibility of hexagon invasion fronts for a non-variational form of the Swift-Hohenberg equation,
\begin{equation}\label{e:non_SH}
u_t = -(1+\Delta)^2u - \mu u + \nu u^2  - u^3 + \beta |\nabla u|^2,
\end{equation}
where $\beta$ is another real parameter. In Figure~\ref{f:non_var_comp}, we plot the compatibility diagram for the $\langle10\rangle$- and $\langle11\rangle$-fronts for $(\mu,\nu,\beta)=(0.15,2,-0.5)$. For these values the existence boundaries of the cellular hexagons are highly complex, and we do not plot them. As with the variational case, we again see that the only compatible hexagon fronts are those that involve perfect hexagons and that the compatible $\langle10\rangle$-front propagates quicker than the compatible $\langle11\rangle$-front. This provides evidence that the conjecture~\ref{c:hex_conjecture} is also true for general pattern forming systems. 

\section{Planar Patch Invasion}\label{s:patch_invasion}
In this section, we will investigate what happens to patches of cellular pattern that invade the trivial state. The conjecture for pattern selection of stationary hexagon patches has already been verified numerically in~\cite{lloyd2008} (see also Figure~\ref{f:stat_patch_select}). Time simulations of the SH equation are carried out on a doubly periodic box of size $[-60\pi,60\pi]^2$ with $2^{10}$ Fourier modes in both $x$ and $y$ and time step $0.01$.

The interfaces associated with either $x=0$ or $y=0$ are tracked by forming the 1D interpolant of the solution $u$ along these lines and looking for when $u=0.5$ using \textsc{Matlab's} \verb1fzero1 routine; see~\cite{lloyd2019} for more details. 

\begin{figure}[htb]
	\centering
	\includegraphics[width=\linewidth]{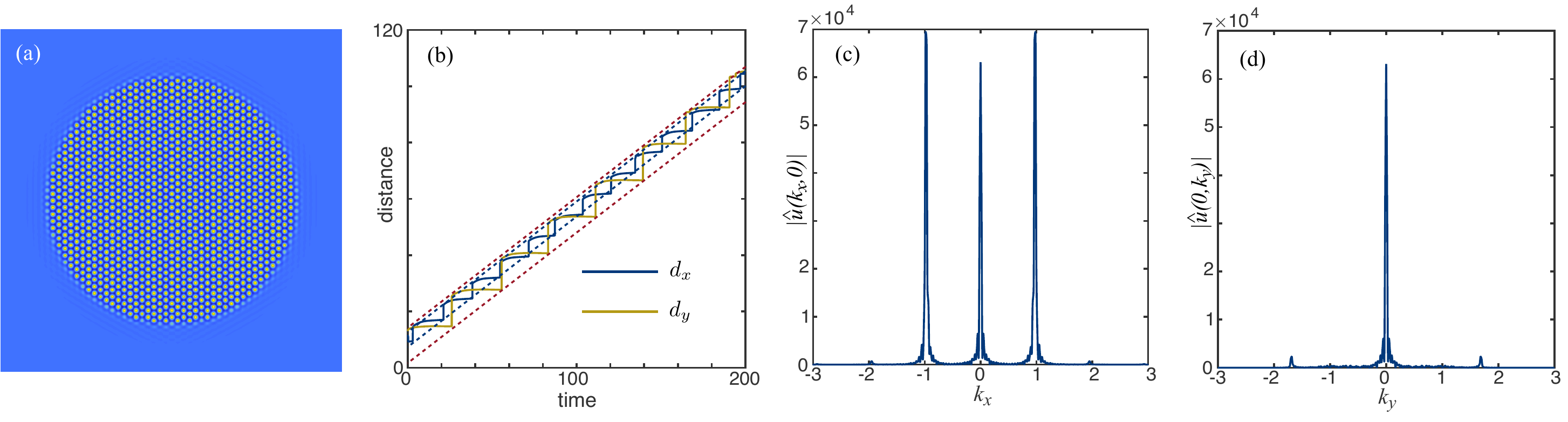}
	\caption{Hexagon patch invasion in the quadratic-cubic SH equation~(\ref{e:sh}) with $(\mu,\nu)=(0.15,1.6)$. Panel (a) shows the patch at $t=200$ and panel (b) shows the interface locations in the $x$- and $y$-directions at the mid-points given by $d_x$ and $d_y$, respectively. The upper blue dashed line is given by a fitted line given by $0.47t+10$, the lower blue dashed line corresponds to a fitted line given by $0.48t+5$, the upper dashed red line is $0.47t+11$ and the lower dashed red line is $0.47t-0.8$. Mean time between jumps in $x$ is $\sim16$ (from $[10]$-front BVP computation $2\pi/0.42\sim 15$) and in $y$ is $\sim 24$ (from $[11]$-front BVP computation $2\pi/0.25\sim 25$). (c) and (d) we plot of the absolute value of the Fourier transform, $\hat u(k_x,0)$ and $\hat u(0,k_y)$ of $u(x,0)$ and $u(0,y)$, respectively. The peak non-trivial Fourier mode occurs at $(k_x,k_y)\sim(0.98,1.68)$ which corresponds to a wavenumber of $(k_x,k_y)\sim(0.49,0.84)$ of the cellular hexagons (compared with BVP compatibility computation where we find that $(k_x,k_y)\sim(0.49,0.84)$).\label{f:hex_patch_speed}}
\end{figure}

We look at hexagon patch invasion in the quadratic-cubic SH equation with $(\mu,\nu)=(0.2,1.6)$ and plot the right most interface distance at $y=0$ which we call $d_x$ and the top most interface distance at $x=0$ which we call $d_y$; see Figure~\ref{f:hex_patch_speed}. We start with the time simulations with a square localized patch of cellular hexagons i.e., 
\[
u(x,y)=\frac{1.2}{6}(\mbox{tanh}(x+d) - \mbox{tanh}(x-d))(\mbox{tanh}(y+d)-\mbox{tanh}(y-d))(e^{ix} + e^{i\frac{x+\sqrt{3}y}{2}}+e^{i(\frac{x-\sqrt{3}y}{2})}+c.c.),
\]
with $d=4\pi$. The growth process is very similar to the snaking behavior of hexagon patches described in~\cite{lloyd2008} where hexagon cells are added first in the middle of any long interface then subsequently additional cells are added along the interface to complete a row. We see at $t=200$, the patch becomes mostly circular and the hexagons appear to be regular rather than stressed. Looking at the distance of the $x$ and $y$ mid-point interfaces, we see that both of these fronts propagate at approximately the same speed (around $0.47$) near to that predicted from the hexagon front compatibility diagram of $0.5$. The mean time between jumps for the $x$-direction interface (corresponding to a $\langle10\rangle$-direction) is $\sim16\approx 2\pi/(ck_x)$ while for the $y$-direction interface (corresponding to a $\langle11\rangle$-direction) it is $\sim24\approx2\pi/(ck_y)$ and the ratio between these mean jumps is approximately $\sqrt{3}$ confirming that the propagation speed is the same. Both of the front interface speeds are close to those predicted from the hexagon front computations. It is hard to detect from the time simulations the prediction from the compatibility diagrams that the $\langle10\rangle$-front should propagate slightly faster than the $\langle11\rangle$-front. However, we find the lower fitted dashed blue curve in Figure~\ref{f:hex_patch_speed} suggests that the $\langle10\rangle$-interface propagates slightly faster corroborating our prediction. This radial growth process appears to be stable. We also compute the Fourier transform, $\hat u(k,0)$, of $u(x,0)$  at $t=200$, and we plot the magnitude of $\hat u(k,0)$ in Figure~\ref{f:hex_patch_speed}(c). We find that the largest Fourier modes occurs at approximately $0.98$ corresponding to a hexagon wavenumber $k_x=0.49$ which compares well with the compatibility prediction of $k_x=0.4857$.  

\begin{figure}[htb]
	\centering
	\includegraphics[width=\linewidth]{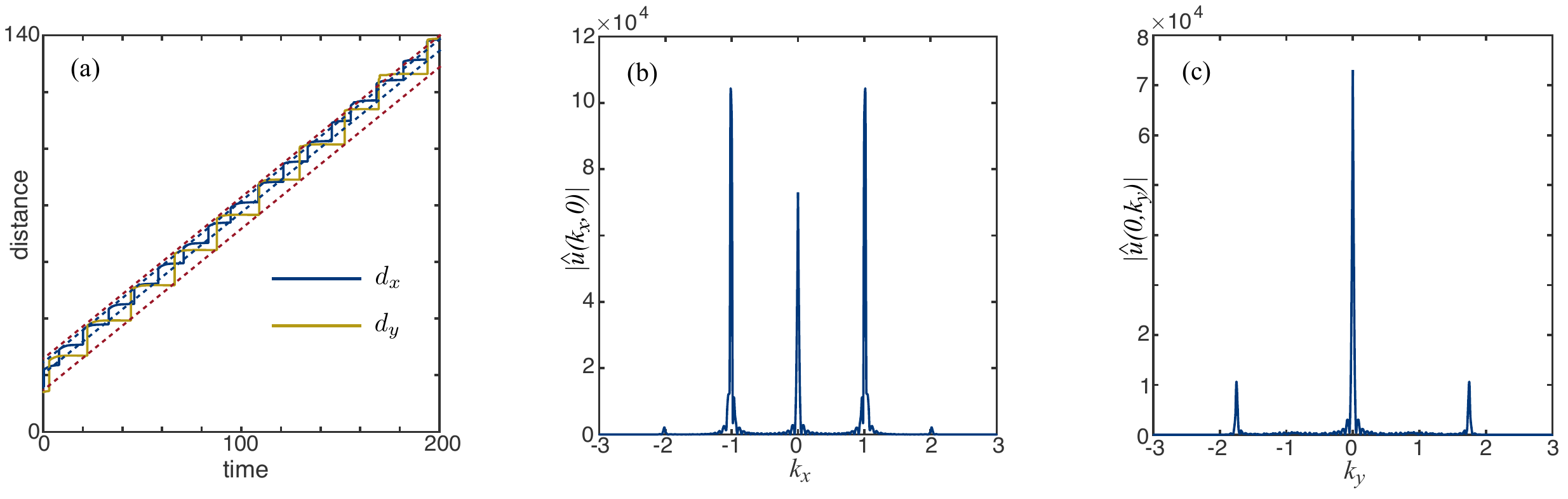}
	\caption{(a) Interface locations of the hexagon patch invasion in the non-variational SH equation~(\ref{e:non_SH}) with $(\mu,\nu,\beta)=(0.15,2,-0.5)$. The upper blue dashed line denotes a fitted line given by $0.59t+18$, the lower blue dashed line is $0.59t+23$, the upper dashed red line is $0.59t+13$ and the lower dashed red line is $0.59t+23$. Mean time between jumps in $x$ is $\sim12$ (from $[10]$-front BVP computation $2\pi/0.54\sim 12$) and in $y$ is $\sim 21$ (from $[11]$-front BVP computation $2\pi/0.31\sim 20$) (b) A plot of the Fourier transform of $u(x,0)$ at $t=200$. The peak non-trivial Fourier mode occurs at $(k_x,k_y)\sim(1.02,1.75)$ which corresponds to a wavenumber of $(k_x,k_y)\sim(0.51,0.875)$ of the cellular hexagons (compared with BVP compatibility computation where we find that $(k_x,k_y)\sim(0.504,0.876)$). 
	\label{f:hex_patch_speed_nonvar}}
\end{figure}

In order to explore the heuristic compatibility condition further, we next carry out time simulations of the non-variational SH equation~(\ref{e:non_SH}) with $(\mu,\nu,\beta)=(0.15,2,-0.5)$. In Figure~\ref{f:hex_patch_speed_nonvar}(a), we plot the growth of a localized hexagon patch to $t=200$. The final state is very similar to Figure~\ref{f:hex_patch_speed}(a) and so we do not plot it. Again, we plot the distance of the $x$ and $y$ mid-point interfaces, and we see that both of these fronts propagate at approximately the same speed (around $0.59$) near to that predicted from the hexagon front compatibility diagram of $\sim0.62$. From the compatibility criterion in Figure~\ref{f:non_var_comp}, we predict that $(k_x,k_y)=(0.5041,0.8760)$ and this compares well with the peak non-trivial Fourier mode of $u(x,0)$ at $t=200$ shown in Figure~\ref{f:hex_patch_speed_nonvar}(b) where we find that the cellular hexagons have a wavenumber of roughly $k_x=0.51$.

\section{Conclusion}\label{s:discussion}

{\bf Summary.} In this paper, we have explored the pattern selection mechanism of both stationary and invading planar, distorted hexagon fronts in the Swift-Hohenberg equation. We find that the widths of the hexagon fronts have a significant effect on the snaking regions. We then introduced the idea of a compatibility diagram to find compatible hexagon fronts that could make up a 2D hexagon patch. It is found that for stationary fronts, only the perfect hexagons or squares are compatible. Numerically, we then explored hexagon invasion fronts using the far-field core decomposition idea described in~\cite{lloyd2019}. We find again that the only stable compatible hexagon fronts are those involving perfect hexagons in their far-field and this is also true if we break the variational structure of the SH equation. It is also found for $\langle10\rangle$-hexagon fronts with small enough $k_y$, that there is a fold instability for small $\mu$ beyond which time simulations show invasion fronts that deposit hexagons with defects. We find no such instability for the $\langle11\rangle$-fronts suggesting they are more robust in producing defect-free patterns in their wake. Finally, we compared our predictions from the compatibility diagrams of the hexagon fronts with hexagon patch growth and find good agreement. 

{\bf Open Problems.} This work raises a series of further questions and problems to look at. Even at a basic level of understanding the spectral properties of stationary cellular distorted hexagons appears to be lacking. On a formal level much progress was made in the 1990s and early 2000s (see for instance~\cite{gunaratne1994,pena2001,hoyle2006}) but there lacks numerical bifurcation algorithms building on the ideas in~\cite{radss,sherratt2012} to trace out stability boundaries with respect to rectangular and rhombic perturbations. Putting the spectral properties on a firmer foundation would also help in understanding the hexagon invasion fronts. 
 
Verifying the compatibility condition for the hexagon fronts involved in the patch invasion process remains a major open problem. 
We highlight that the numerical methods described in this paper and~\cite{lloyd2019} do not depend on the SH structure and are generally applicable to reaction-diffusion systems, nonlocal equations e.g.,~\cite{alnahdi2018,ly2019,ophaus2018,rankin2014} and 3D patterns~\cite{uecker2019} allowing one to compute the far-field wavenumber selection of both stationary and propagating cellular pattern fronts. 

It would be interesting to explore the emergence of defects in the $\langle10\rangle$-front in the amplitude equations. Given that the computations of the hexagon invasion fronts involve large discretizations, development of time-stepper numerical continuation algorithms (see for instance~\cite{sanchez2004}) and suitable preconditioners to solve the linear step in the Newton solver appear to be crucial in continuing the time-periodic defects shown in Figure~\ref{f:stat_hex_snake}(d). In terms of further work, propagating fronts connecting distorted hexagons to stripes or other distorted hexagons, penta-hepta defects etc. would be another interesting avenue to explore~\cite{subramanian2020}; see for instance~\cite{wetzel2018} for various stationary fronts between hexagons and stripes. The pattern selection mechanism in these fronts has yet to be explored even in the stationary case.

\section*{Code availability}
Codes used to produce the results in this paper are available at:

\url{https://github.com/David-JB-Lloyd-Lab/Invasion-fronts}

\section*{Acknowledgments}
DJBL would like to thank  Daniele Avitabile, Cedric Beaume, Jonathan Dawes, Daniel Hill, Edgar Knobloch, Bj\"{o}rn Sandstede and Arnd Scheel for helpful discussions on this work. We also thank the referees for their helpful comments. 

\section*{Appendix}
In this appendix, we briefly discuss the bordering method we employ to solve the linear system involved in carrying out a Newton step. 

We follow the method described in~\cite{phipps2006}. For solving the Newton step for parameter continuation of the discretization of (\ref{e:PDE_w}) and (\ref{e:phase_cond}), we need to solve a linear system of the form
\begin{equation}\label{e:bordered}
\left[\begin{array}{cc}
\mathbf{J} & \mathbf{A}\\
\mathbf{B}^T & \mathbf{C}
\end{array} \right]\left[\begin{array}{c}
\mathbf{X}\\\mathbf{Y}
\end{array} \right] = \left[\begin{array}{c}
\mathbf{F}\\\mathbf{G}
\end{array} \right],
\end{equation}
where $\mathbf{J}\in\mathbb{R}^{n\times n}$ ($n$ is the number of spatial discretization points), $\mathbf{A}\in\mathbb{R}^{n\times m}$ ($m=3$ is the number of phase conditions -- two from (\ref{e:phase_cond}) and one from the pseudo-arclength continuation), $\mathbf{B}^T\in\mathbb{R}^{m\times n}$, and $\mathbf{C}\in\mathbb{R}^{m\times m}$. 

In order to solve this system,  Phipps and Salinger~\cite{phipps2006} suggest an extension to the Householder pseudo-arclength technique by Walker~\cite{walker1999}. First the bordered system~(\ref{e:bordered}) is rearranged into the form
\begin{equation}\label{e:bordered2}
\left[\begin{array}{cc}
\mathbf{C} & \mathbf{B}^T\\
\mathbf{A} & \mathbf{J}
\end{array} \right]\left[\begin{array}{c}
\mathbf{Y}\\\mathbf{X}
\end{array} \right] = \left[\begin{array}{c}
\mathbf{G}\\\mathbf{F}
\end{array} \right],
\end{equation}
and then we carry out a QR factorization of
\[
\left[\begin{array}{c}
\mathbf{C}^T\\
\mathbf{B}
\end{array} \right] = \mathbf{Q}\left[\begin{array}{c}
\mathbf{R}\\\mathbf{0}
\end{array} \right].
\]
We then introduce
\begin{equation}\label{e:ZYZX}
\left[\begin{array}{c}
\mathbf{Z_Y}\\
\mathbf{Z_X}
\end{array} \right] = \mathbf{Q^T}\left[\begin{array}{c}
\mathbf{Y}\\
\mathbf{X}
\end{array} \right] ,
\end{equation}
and the bordered system (\ref{e:bordered2}) can be written as
\[
\left[
\begin{array}{cc}
\mathbf{R^T} & \mathbf{0}\\
\mathbf{[A} & \mathbf{J]Q}
\end{array}
\right]\left[\begin{array}{c}
\mathbf{Z_Y}\\
\mathbf{Z_X}
\end{array} \right] = \left[\begin{array}{c}
\mathbf{G}\\\mathbf{F}
\end{array} \right],
\]
where $[\mathbf{A}\;\; \mathbf{J}]$ denotes the augmented matrix of $\mathbf{A}$ and $\mathbf{J}$. Hence, we find that,
\begin{equation}\label{e:PZX}
\mathbf{Z_Y} = \mathbf{R^{-T}G},\qquad \mathbf{PZ_X} = \mathbf{\tilde F},
\end{equation}
where 
\[
\mathbf{PZ_X} = \mathbf{[A\;\; J]Q}\left[\begin{array}{c}
\mathbf{0}\\
\mathbf{Z_X}
\end{array} \right],\qquad \mathbf{\tilde F} = \mathbf{F - [A\;\; J]Q}\left[\begin{array}{c}
\mathbf{Z_Y}\\
\mathbf{0}
\end{array} \right].
\]
The solution to the bordered system~(\ref{e:bordered2}) is found by solving (\ref{e:PZX}) for $\mathbf{Z_X}$ and $\mathbf{Z_Y}$ and inverting (\ref{e:ZYZX}) to find,
\[
\left[\begin{array}{c}
\mathbf{Y}\\
\mathbf{X}
\end{array} \right] = \mathbf{Q}\left[\begin{array}{c}
\mathbf{Z_Y}\\
\mathbf{Z_X}
\end{array} \right].
\]

In order to solve the second equation in (\ref{e:PZX}), we show that $\mathbf{P}$ can be written as a rank $m$ update to $\mathbf{J}$. The matrix $\mathbf{Q}$ can be written as the product of Householder matrices (of the form $\mathbf{H}=\mathbf{I}-2\mathbf{vv^T}$) and has a ``$WY$ representation'' of the form $\mathbf{Q}=\mathbf{I}+\mathbf{VW^T}$, where $\mathbf{V},\mathbf{W}\in\mathbb{R}^{n\times m}$. Such a representation of $\mathbf{Q}$ is memory storage intensive but Schreiber and Loan~\cite{schreiber1989} introduced a memory storage efficient ``compact $WY$ representation'' of $\mathbf{Q}$ by noting that if $\mathbf{Q}=\mathbf{I} + \mathbf{WTW^T}$, where $\mathbf{W}\in\mathbb{R}^{n\times j}$ and $\mathbf{T}\in\mathbb{R}^{j\times j}$ is upper triangular, and $\mathbf{H}=\mathbf{I}-2\mathbf{vv^T}$, with $\mathbf{v}\in\mathbb{R}^n$, then the product $\mathbf{QH}$ could also be written in the form $\mathbf{I}+\mathbf{\widetilde{W}\widetilde{T}\widetilde{W}^T}$ for some new matrices $\mathbf{\widetilde{W}}\in\mathbf{R}^{n\times(j+1)}$ and $\mathbf{\widetilde{T}}\in\mathbf{R}^{(j+1)\times(j+1)}$ is upper triangular. 

Using the compact $WY$ representation for $\mathbf{Q}$ such that $\mathbf{Q} = \mathbf{I} + \mathbf{WTW^T}$, where $\mathbf{W}\in\mathbb{R}^{n\times m}$ and $\mathbf{T}\in\mathbb{R}^{m\times m}$ is upper triangular, we split $\mathbf{W = [W_1\;\; W_2]^T}$ where $\mathbf{W_1}\in\mathbb{R}^{m\times m}$ and $\mathbf{W_2}\in\mathbb{R}^{n\times m}$, then $\mathbf{P}$ can be written as a rank $m$ update to $\mathbf{J}$
\begin{equation}\label{e:P}
\mathbf{P = J + \left((AW_1 + JW_2)T\right)W_2^T =: J + UV^T}.
\end{equation}
The matrix $\mathbf{P}$ is expected to be a non-singular (away from any bifurcations), full matrix that can be solved either iteratively (where Phipps~\cite{phipps2006} suggest using a preconditioner which includes $\mathbf{UV^T}$ terms for non-zero entries of $\mathbf{J}$) or using a form of the Woodbury matrix formula for a singular matrix $\mathbf{J}$; see Riedel~\cite{riedel1992}. In practice, we have found that it is sufficient (and efficient/fast) to use the standard Woodbury matrix formula
\[
\mathbf{(J + UV^T)^{-1} = J^{-1} - J^{-1}U(I + V^TJU)^{-1}V^TJ^{-1}}.
\]
Then $\mathbf{Z_X}$ is found by solving two linear systems
\[
\mathbf{JZ = U},\qquad \mathbf{JY = \tilde F},
\]
to yield $\mathbf{Z_X = Y - Z(I+V^TZ)^{-1}V^TY}$. The two linear systems are solved by carrying out a sparse LU decomposition of $\mathbf{J}$ using \textsc{Matlab's} \verb1decomposition1 routine.

\bibliographystyle{siam}
\bibliography{swift_depinning_bib}

\end{document}